\documentclass{aa}
\pdfoutput=1  % to force arXiv to process this file using pdflatex

\usepackage{natbib}
\usepackage[]{graphicx}
\usepackage{txfonts}

\usepackage{color} % to mark changes in revision with color
\usepackage[normalem]{ulem}  % to use \sout for striketrough
\def\pow#1#2{#1$\times$10$^{#2}$}
\def\micron{$\mu$m}
\def\kms{$\mathrm{km}\,\mathrm{s}^{-1}$}  % km/s
\def\Kkms{$\mathrm{K\,km}\,\mathrm{s}^{-1}$}  % K km/s

\def\pccm{$\mathrm{cm}^{-3}$}
\def\Tmb{$T_\mathrm{mb}$}  % main beam temperature
\def\intintens{$\int T_\mathrm{mb} \mathrm{d}V$}
\def\vlsr{$V_\mathrm{lsr}$}  % local standard of rest velocity

\def\HH{H$_2$}
\def\HCOplus{HCO$^+$}
\def\HthCOplus{H$^{13}$CO$^+$}
\def\HHCO{H$_2$CO}
\def\NtwoHplus{N$_2$H$^+$}
\def\thCO{$^{13}$CO}
\def\CstO{C$^{17}$O}
\def\SOtwo{SO$_2$}

\def\methanol{CH$_3$OH}

\def\CtfS{C$^{34}$S}
\def\CCH{C$_2$H}

\def\tSLSchempaper{Van der Wiel et al.~(2011b, in prep.)}

\def\pCHESSAFGL{(Van der Wiel et al.~2011a, in prep.)}

\def\markchanges{yes}  % yes or no

\def\marked{yes}
\def\unmarked{no}

\definecolor{orange}{rgb}{1.0,0.15,0}
\definecolor{darkgreen}{rgb}{0.0,0.0,0.4}

\ifx\markchanges\marked % mark changes:
	\newcommand{\removed}[1]{\textcolor{orange}{[}\sout{#1}\textcolor{orange}{]}}  % mark removed stuff
	
\fi
\ifx\markchanges\unmarked % DO NOT mark changes:
	\newcommand{\removed}[1]{}  % do not show removed parts
	
\fi

\begin{document}

\title{The JCMT Spectral Legacy Survey: physical structure of the molecular envelope of the high-mass protostar AFGL2591}
\titlerunning{JCMT-SLS: The molecular envelope of AFGL2591}

\author{M.~H.~D.~van der Wiel\inst{1,2} \and F.~F.~S.~van der Tak\inst{2,1} \and M.~Spaans\inst{1} \and G.~A.~Fuller\inst{3} \and R.~Plume\inst{4} \and H.~Roberts\inst{5} \and J.~L.~Williams\inst{3}}

\institute{
% 1: Kapteyn Groningen
Kapteyn Astronomical Institute, University of Groningen, P.O.~Box 800, 9700\,AV, Groningen, The Netherlands\\email: \texttt{wiel@astro.rug.nl}
\and 
% 2: SRON Groningen
SRON Netherlands Institute for Space Research, P.O. Box 800, 9700\,AV, Groningen, The Netherlands
\and
% 3: University of Manchester, UK
Jodrell Bank Centre for Astrophysics, Alan Turing Building, University of Manchester, Manchester, M13 9PL, UK
\and
% 4: U. of Calgary, Canada
Department of Physics and Astronomy, University of Calgary, Calgary, T2N 1N4, AB, Canada
 \and
% 5: Queen's Belfast, Northern Ireland
Astrophysics Research Centre, School of Mathematics and Physics, Queen's University of Belfast, Belfast, BT7 1NN, UK
}

\date{Received November 5, 2010 / Accepted January 27, 2011}

\abstract 
% context:
{The understanding of the formation process of massive stars ($\ga$8\,$M_\odot$) is limited, due to a combination of theoretical complications and observational challenges. The high UV luminosities of massive stars give rise to chemical complexity in their natal molecular clouds, and affect the dynamical properties of their circumstellar envelopes.} 
% aims:
{We investigate the physical structure of the large-scale ($\sim$$10^4$--$10^5$\,AU) molecular envelope of the high-mass protostar AFGL2591. }
% methods:
{Observational constraints are provided by spectral imaging in the 330--373\,GHz regime from the JCMT Spectral Legacy Survey and its high frequency extension. While the majority of the $\sim$$160$ spectral features from the survey cube are spatially unresolved, this paper uses the 35 that are significantly extended in the spatial directions. For these features we present integrated intensity maps and velocity maps. 
The observed spatial distributions of a selection of six species are compared with radiative transfer models based on (i) a static spherically symmetric structure, (ii) a dynamic spherical structure, and (iii) a static flattened structure. }
% results:
{
% observational results:
The maps of CO and its isotopic variations exhibit elongated geometries on scales of $\sim$$100$\arcsec, and smaller scale substructure is found in maps of \NtwoHplus, o-\HHCO, CS, \SOtwo, \CCH, and various \methanol\ lines. In addition, a line of sight velocity gradient is apparent in maps of all molecular lines presented here, except SO, \SOtwo, and \HHCO. We find two emission peaks in warm ($E_\mathrm{up} \sim 200$\,K) \methanol\ separated by 12\arcsec\ (12\,000\,AU), indicative of a secondary heating source in the envelope. 
% modeling results:
The spherical models are able to explain the distribution of emission for the optically thin \HthCOplus\ and \CtfS, but not for the optically thick HCN, \HCOplus, and CS, nor for the optically thin \CstO. The introduction of velocity structure mitigates the optical depth effects, but does not fully explain the observations, especially in the spectral dimension. A static flattened envelope viewed at a small inclination angle does slightly better. 
}
% conclusions:
{
Based on radiative transfer modeling, we conclude that a geometry of the envelope other than an isotropic static sphere is needed to circumvent line optical depth effects. 
We propose that this could be achieved in circumstellar envelope models with an outflow cavity and/or inhomogeneous structure at scales $\lesssim 10^4$\,AU. The picture of inhomogeneity is supported by observed substructure in at least six different species. 
}

\keywords{stars: formation -- ISM: molecules -- ISM: individual objects: AFGL2591 -- Techniques: imaging spectroscopy}

\maketitle

% ---------- INTRODUCTION ---------
\section{Introduction}

% -------- star formation in general
A collapsing protostellar cloud is heated by gravitational contraction of the envelope and radiation from the central protostar. Therefore, the process of star formation depends crucially on the disposal of thermal energy through (molecular) transition lines. Molecular signatures from star-forming regions are an important probe in studies of star formation. 
% -------- The challenges of studying of high-mass star formation
While the energy budget is reasonably well understood for protostars with masses $\lesssim 8\,M_\odot$ \citep[e.g.,][]{evans1999,larson2003}, the problem is more complex for high-mass stars ($\gtrsim 8\,M_\odot$). Not only do they require more material and involve (orders of magnitude) more energy, high-mass stars form in more deeply embedded natal clouds, and generally in more clustered environments. In addition, high-mass protostars are rarely observed due to the fact that they form less often, and that if they do, their formation timescales are much shorter than for their low-mass counterparts (see \citealt{zinnecker2007} for a review). In addition, young high-mass stars have the ability to significantly alter the chemical composition of their parental clouds by the copious amounts of UV-radiation they emit \citep{stauber2004}. It is exactly this impact that high-mass stars have on the surrounding interstellar medium -- both when they form and when they explode as supernovae -- which makes it important to study their formation process in a broader context.

% -------- what we don't understand about high-mass star formation
Due to the above challenges in the study of high-mass star formation, it is as yet unclear whether a high-mass star forms following a process similar to low-mass star formation (core contraction $\rightarrow$ envelope collapse $\rightarrow$ disk accretion $\rightarrow$ disk evaporation), or that other processes such as competitive accretion or stellar mergers are at play \citep[e.g.,][]{bonnell2001a,bonnellbate2005,hocuk2010b}. In any case, much higher accretion rates and more energetic feedback processes must be involved in high-mass star formation, providing additional challenges in theoretical modeling. 
% -------- why study envelopes of massive YSOs
Investigations into the physical and chemical structure (at various scales) of the molecular envelopes of massive young stars can shed light on the accretion process and the energy distribution and dissipation in regions of high-mass star formation \citep{evans1999,cesaroni2005}. One hypothesis to consider is whether high-mass star formation in its first stages proceeds less isotropically than low-mass star formation, undergoing non-spherical accretion even before material reaches a potential circumstellar disk.

% --------The massive YSO AFGL2591 
This paper studies the molecular envelope of \object{AFGL2591}, an isolated massive star-forming region located in the Cygnus-X region at galactic coordinates $(\ell,b)=(78\fdg9, 0\fdg71)$. The distance to AFGL2591 is between 0.5 and 2.0\,kpc \citep[see][]{vandertak1999,schneider2006}. While the commonly assumed average distance for the Cygnus-X complex is 1.7\,kpc \citep{motte2007}, the distance to individual objects in this local spiral arm structure varies. For consistency with earlier work on this source, we adopt a distance of 1.0\,kpc and we refer to discussion in \citet{vandertak1999} concerning the effects of a larger assumed source distance. The protostar is estimated by \citet{vandertak2005} to have a mass of $16\,M_\odot$. The object emits most of its energy in the infrared, due to re-radiation of optical/UV radiation from the central object by the surrounding gas and dust. It has a bolometric luminosity of $\sim$\pow{2}{4}\,$L_\odot$ (at 1\,kpc) and exhibits powerful large ($> 1 \arcmin$) bipolar molecular outflows \citep{lada1984,hasegawa1995}. 

% --------structure of its molecular envelope: usually assumed spherical, sometimes with cavity
Previous modeling of the molecular envelope of AFGL2591 has assumed a spherical morphology and found that observations of molecular lines and dust can be explained using a power law density profile \citep{vandertak1999,vandertak2000jul,doty2002,stauber2005,benz2007,dewit2009}. \citet{vandertak1999} suggested the presence of cavities in the envelope, carved out by the outflows, as a solution to optical depth effects which prevent molecular line radiation from escaping the cloud if the envelope is spherical and isotropic. After evidence of such cavities was observed at high spatial resolution ($\sim2\arcsec$) in the near-infrared by \citet{preibisch2003}, the idea was invoked again by \citet{bruderer2010a,bruderer2010b} to allow far-UV radiation from the protostar to directly excite molecular species in the envelope. The overall structure of the envelope, however, is still assumed to be spherical.

% --------setup of this paper
This paper is structured as follows. In Sect.~\ref{sec:obs} we describe the unbiased sub-millimeter spectral imaging survey of the molecular envelope of AFGL2591; the resulting maps of integrated intensity and velocity structure for 35 molecular transitions are presented in Sect.~\ref{sec:results_spatialextent}. While this paper focuses on the spatial domain in order to investigate the large-scale ($>10^4$\,AU) envelope, a full exploitation of the spectral domain will be addressed in \tSLSchempaper. Sect.~\ref{sec:modeling} describes various approaches to modeling a selection of six molecular emission lines using a radiative transfer code. Section~\ref{sec:discussion}, finally, is devoted to the discussion of the results of the radiative transfer modeling, a phenomenological interpretation of the lines that are not modeled, and general conclusions.

% ---------- OBSERVATIONS ---------
\section{Observations}
\label{sec:obs}

\subsection{HARP-B and the JCMT Spectral Legacy Survey}

% == HARP-B ==
The observations presented here were taken with the 16-element Heterodyne Array Receiver Programme B (HARP-B) and the Auto-Correlation Spectral Imaging System (ACSIS) correlator \citep{smith2008,dent2000,buckle2009} at the James Clerk Maxwell Telescope\footnote{The James Clerk Maxwell Telescope is operated by the Joint Astronomy Centre on behalf of the Science and Technology Facilities Council of the United Kingdom, the Netherlands Organisation for Scientific Research, and the National Research Council of Canada.} (JCMT) on Mauna Kea, Hawai'i. The strength of the multipixel HARP-B instrument lies in the combination of high-resolution (1\,MHz, $\approx$1\,\kms) heterodyne spectroscopy with instantaneous mapping capabilities. 

% == JCMT-SLS and high-frequency extension ==
The JCMT Spectral Legacy Survey \citep[SLS,][]{plume2007} has been allocated a total of 456 hours to map four Galactic targets in the spectral region between 330 and 360\,GHz. The targets are: the Orion Bar, a prototypical photodissociation region \citep{vanderwiel2009}; NGC1333 IRAS 4, a low-mass star-forming region; W49A, an active star formation complex \citep{roberts2011}; and the isolated high-mass star-forming region AFGL2591, which is the topic of this study. The SLS is complemented at higher frequencies (360--373\,GHz) by HARP maps from separate proposals between 2007 and 2010. 
% observing dates:
Observations for the SLS (330--360\,GHz) started immediately after the commissioning of the HARP-B instrument; AFGL2591 was observed in November 2007, March, July and September 2008, May--July 2009, and May--July 2010. The high frequency maps were observed in August and September 2007 and July 2008. AFGL2591 is the first of the four SLS targets for which the spectral imaging has been completed in the 330--360 and 360--373\,GHz spectral regions.

% == HARP observing mode, weather conditions ==
The 16 receptors of the HARP receiver are distributed in a 4$\times$4 pattern, regularly spaced at 30$\arcsec$ intervals. The \texttt{harp4\_mc} jiggle position switch mode was used to create maps of a 2\arcmin\,$\times$\,2\arcmin\ area with pixels spaced by 7.5$\arcsec$, roughly half of the JCMT beam width at the relevant frequencies (15$\arcsec$ at 345\,GHz). 
While the SLS observations were done in relatively wet weather at $0.12<\tau_{225\,\mathrm{GHz}}<0.20$, the observations above 360\,GHz require $\tau_{225\,\mathrm{GHz}}<0.08$, since the Earth's atmosphere becomes increasingly opaque as the frequency approaches 375\,GHz. Apart from the weather constraints, the observing strategy and reduction method is identical for the high frequency data and the SLS data. 

The spectra of AFGL2591 are calibrated by measurements at an off position 340$\arcsec$ east of the target. After inspection of the calibrated data product, we note that there is a possibility of $^{12}$CO~3--2 contaminating emission at the off position. There is no evidence of contamination from the off position for other spectral lines. Redundancy for bad receptors is built in by consecutive observations at an offset of 15$\arcsec$, or by rotation of the entire array during periods where more than three of the sixteen receptors were not functioning. Redundancy for frequency spikes is provided by offsetting the central frequency of consecutive observations, each spanning 1\,GHz bandwidth, by 200 or 400\,MHz. Pointing was checked regularly and corrections were generally $<2$--$3\arcsec$.

\subsection{HARP-B data processing}
% == data reduction ==
The raw time series files are processed in blocks of 1\,GHz bandwidth, using standard procedures from the \texttt{Starlink} software package. Each scan file is inspected for bad receptors, baseline issues and frequency spikes, which are masked before scan files at the same frequency are summed and converted from time series into three-dimensional data cubes (RA$\times$Dec$\times$frequency). The generally noisy band edges are discarded, typically leaving 0.82\,GHz of usable bandwidth per block. Baselines are fitted to line-free regions in every cube using a spline algorithm; the generally nearly linear baselines are subsequently subtracted. The baseline corrected cubes are mosaicked into a larger cube spanning (2\arcmin)$^2$ on the sky and 330.0--373.3\,GHz in frequency. The noise level is reduced by exploiting the frequency overlap between subsequent frequency settings. The spectrum with native frequency bins of 1\,MHz (0.8--0.9\,\kms\ between 373 and 330\,GHz) is resampled to uniform 1\,\kms\ velocity resolution. Intensities are corrected for the JCMT main beam efficiency of 0.61$\pm$0.03 \citep{buckle2009} in the 345\,GHz window and are therefore given in terms of main beam temperature \Tmb. To convert observed ${T_A}^*$ values to flux density units (Jy), multiplication by a factor 29.4 would be necessary.

% figure measured noise
\begin{figure}
  \centering
  \resizebox{\hsize}{!}{\includegraphics{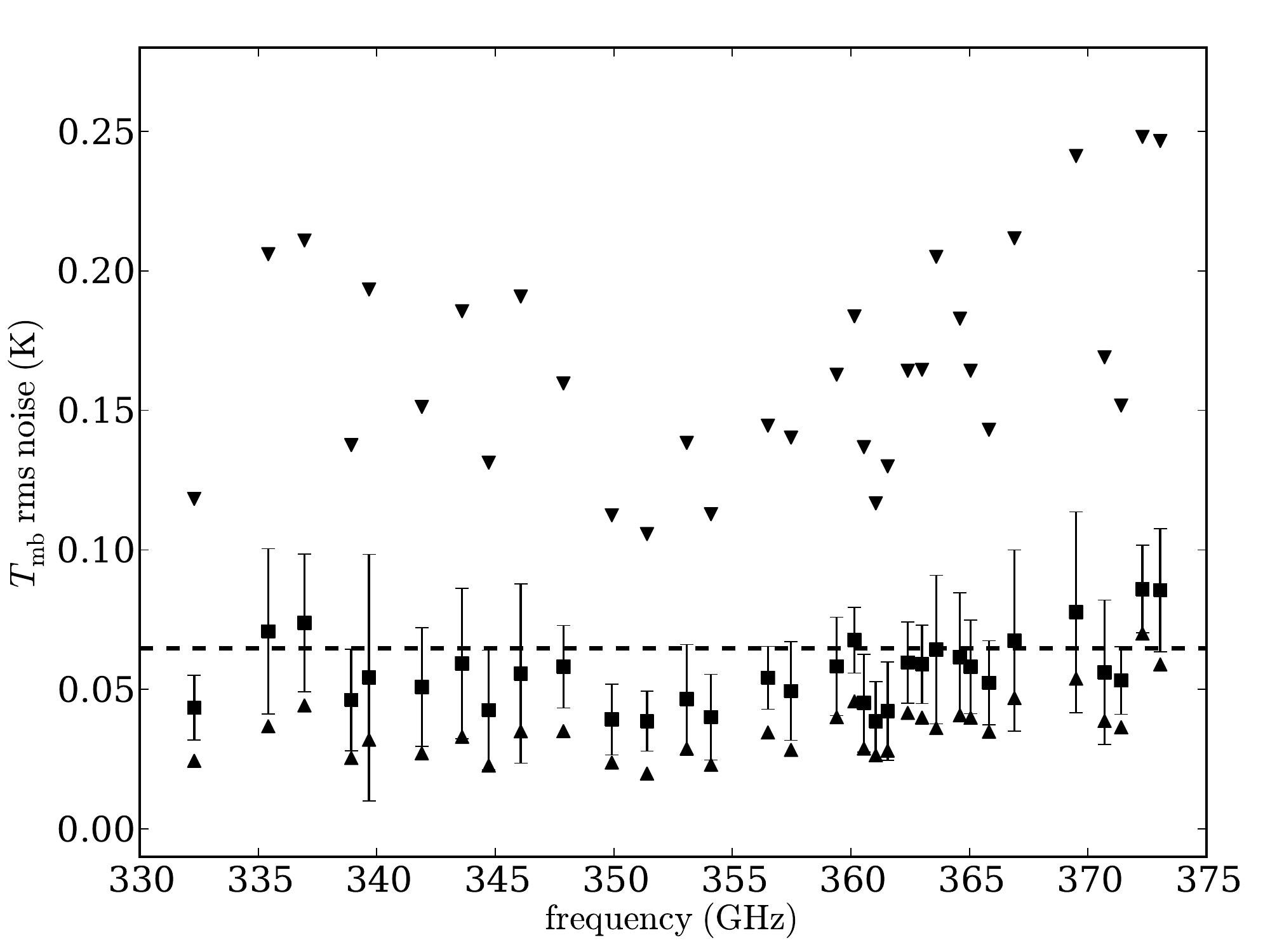}}
  \caption{Root-mean-square (rms) noise statistics of the data product, calculated in 1\,\kms\ channels. The square symbols denote the median rms in all map pixels over a selected frequency range, with their error bars indicating the rms of the rms noise across all map pixels in the same frequency range. The triangle symbols mark the rms value in the least noisy (pointing up) and most noisy pixel (pointing down) in the map. The horizontal dashed line indicates the survey target noise for this object.  }
  \label{fig:noisestats}
\end{figure}

% == noise statistics ==
Figure~\ref{fig:noisestats} shows the median of the root-mean-square (rms) noise in the resulting data cube in some selected line-free frequency ranges, as well as the spread in rms across the map pixels. The noise in \Tmb\ units in both the 330--360\,GHz and the 360--373\,GHz range are generally close to the \Tmb\ target noise level of 65\,mK (converted from 25\,mK in 2.5\,\kms\ channels in ${T_A}^*$ units). The lowest noise values are usually found near the spatial center of the cube.

\subsection{Submillimeter continuum images}

The analysis in this paper uses subsidiary submillimeter continuum images of AFGL2591. These 450 and 850\,\micron\ images have previously been presented by \citet{vandertak2000jul}. The raw images, available from the JCMT archive at the Canadian Astronomy Data Centre, are processed using the standard ORAC-DR pipeline for SCUBA jiggle-maps. An extinction correction is applied, based on sky-dip measurements of the sky optical depth. The resulting maps have units of mJy\,beam$^{-1}$ and are sampled in 3\arcsec\ pixels. The FWHM beam size is 8\arcsec\ for the 450\,\micron\ map, and 14\arcsec\ for the 850\,\micron\ map. This makes the latter map especially suitable for comparison with our HARP-B spectral maps.

%%%%
% Integrated intensity maps
%%%%

\begin{figure*}
  \centering
  \resizebox{\hsize}{!}{\includegraphics{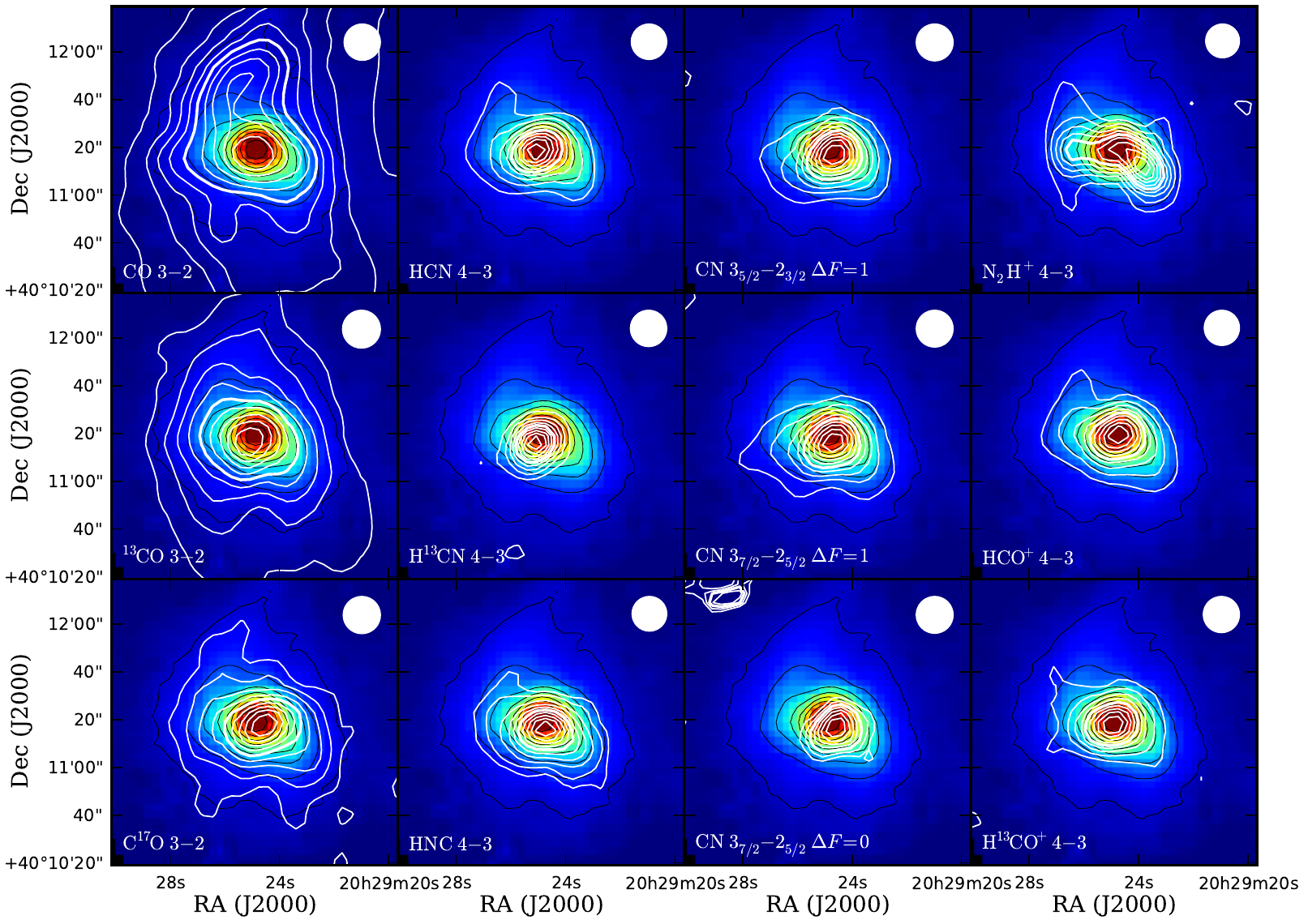}}
  \caption{Maps of emission from the 3--2 transitions of CO, \thCO, \CstO, transitions of N-bearing molecules, \HCOplus\ and \HthCOplus~4--3. White contours represent integrated line emission and are at 90\%, 80\%, \ldots, 10\% of the maximum intensity, listed in Table~\ref{t:maxint}. The contour at 50\% is thicker. An extra contour is added for HNC 4--3 at 5\% of the maximum intensity. The beam size at the relevant frequency is indicated in the top right corner of each panel. To avoid noise in the figures, contours below 0.7\,\Kkms are not shown. SCUBA 850\,\micron\ continuum emission is shown in color scale and thin black contours; the color scale is stretched from 10 to 5600\,mJy\,beam$^{-1}$.
  % (continuum images from Jennifer Williams, priv.~comm.)
  }
  \label{fig:mapsCO_N_HCOplus}
\end{figure*}

\begin{figure*}
  \centering
  \resizebox{0.88\hsize}{!}{\includegraphics{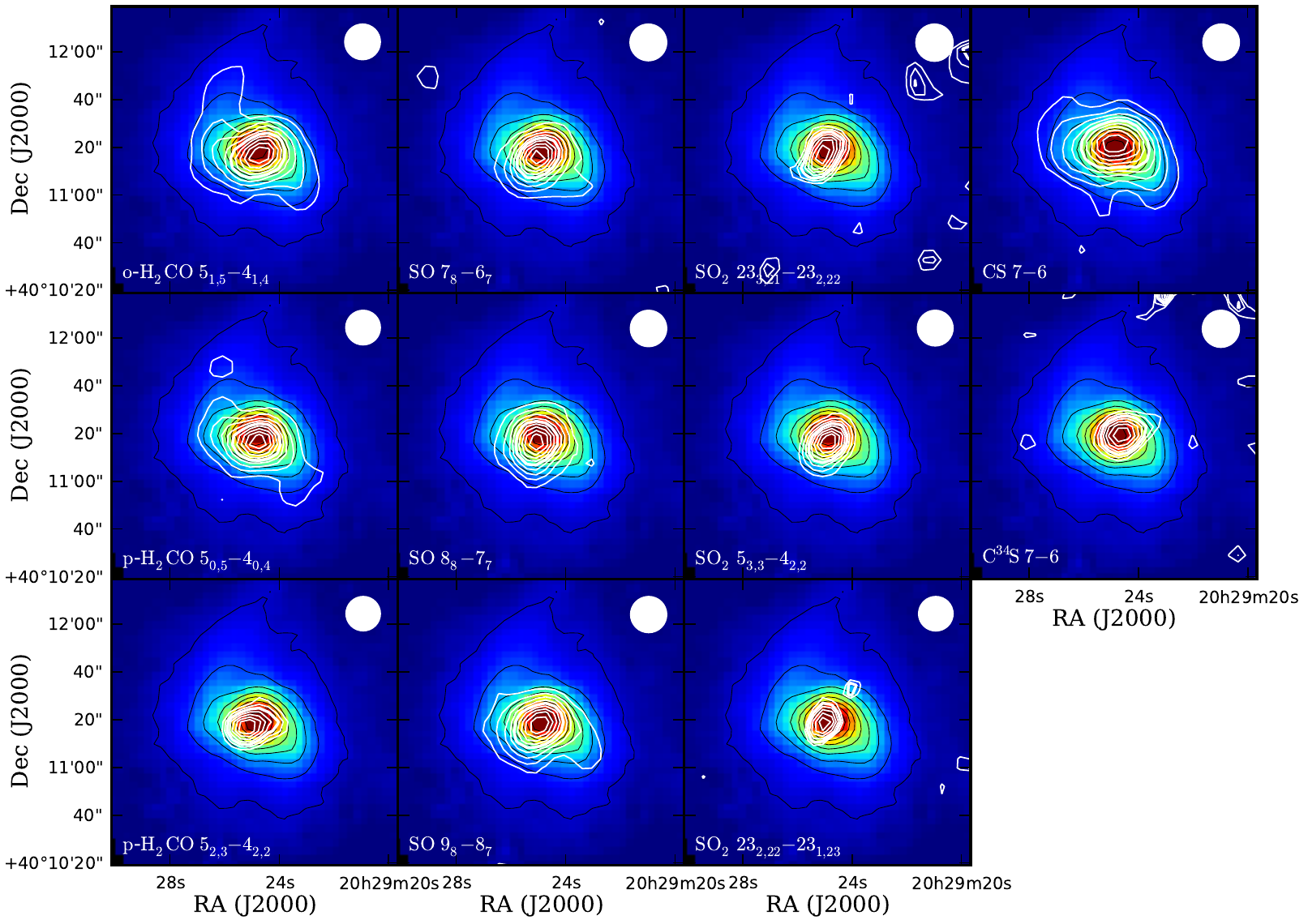}}
  \caption{Like Fig.~\ref{fig:mapsCO_N_HCOplus}, but for \HHCO, SO, \SOtwo, and CS. Maximum intensities are listed in Table~\ref{t:maxint}.} 
  \label{fig:mapsH2CO_SOx_CS}
\end{figure*}

\begin{figure*}
  \centering
  \resizebox{0.88\hsize}{!}{\includegraphics{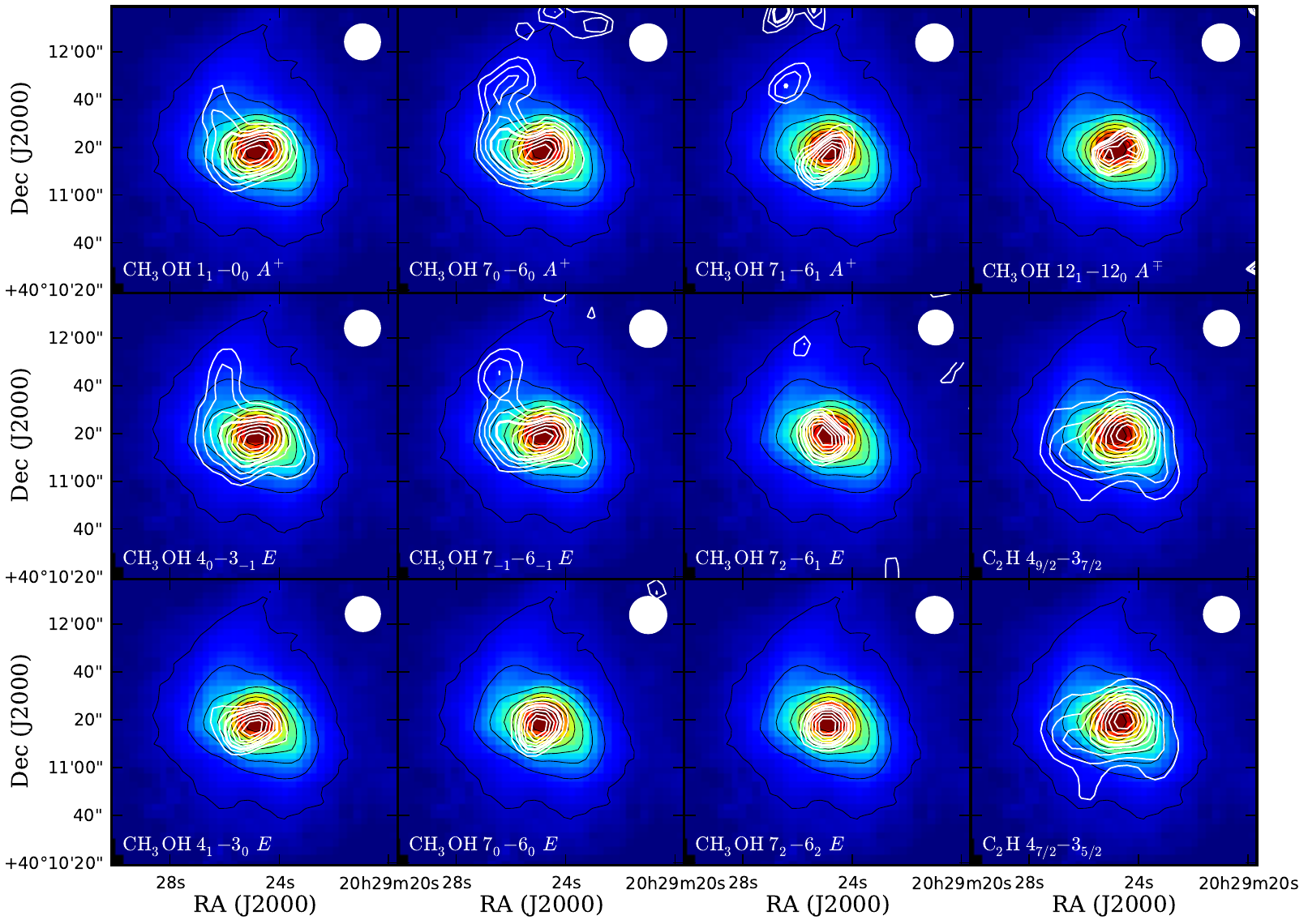}}
  \caption{Like Fig.~\ref{fig:mapsCO_N_HCOplus}, but for \methanol\ and \CCH. Maximum intensities are listed in Table~\ref{t:maxint}. } 
  \label{fig:mapsCH3OH_C2H}
\end{figure*}

% ---------- RESULTS ---------
\section{Observed spatial distribution}
\label{sec:results_spatialextent}

% -------- selection and presentation of extended lines --------
\subsection{Selection of spatially resolved maps}
\label{sec:selectextended}

% selection of spatially extended lines
While the majority of the $\sim$160 spectral lines show a spatial distribution that is confined to the beam size of the JCMT (14--15$\arcsec$ FWHM at the relevant frequency), in this paper we select the lines that are spatially resolved. The selection is made by fitting a two-dimensional Gaussian profile to a spatially resampled (3\farcs75 pixels) integrated line intensity (\intintens) map for each spectral feature. Features are registered in Table~\ref{t:spatialextent} if $a_\mathrm{FWHM} > B_\mathrm{FWHM} + 3 \sigma_a$, with $a_\mathrm{FWHM}$ and $\sigma_a$ the fitted long axis and associated error, and $B_\mathrm{FWHM}$ the JCMT beam size. Spatial extent parameters in Table~\ref{t:spatialextent} are given as measured in the observed maps, without beam deconvolution. We select a total of 35 spectral features with this method. 

% line ID
The association of spectral features to molecular transitions is done using CDMS \citep{muller2005}. In cases where multiple molecular transitions fall within a few MHz of a measured line frequency, we choose the simplest molecule and/or the lowest upper level energy. For the 35 lines selected above, which are relatively strong (\intintens~$\gtrsim 2$\,\Kkms), the identifications listed in Table~\ref{t:spatialextent} are unambiguous. Some features are marked as a blend of several (hyperfine) transitions of the same or different molecules.

\begin{figure}
  \centering
% resize to fraction of hsize to match size of other figure panels
  \resizebox{0.85\hsize}{!}{\includegraphics{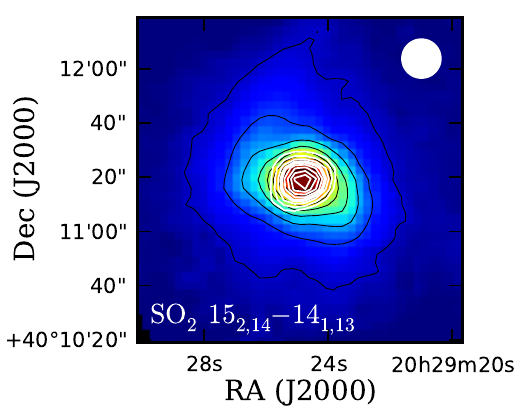}}
  \caption{Spatial distribution of integrated \SOtwo\ $15_{2,14}$--$14_{1,13}$ emission (white contours), which is not extended. The colorscale and black contours represent 850\,\micron\ continuum, like in Fig.~\ref{fig:mapsCO_N_HCOplus}. }
  \label{fig:notextended}
\end{figure}

% description of integrated intensity maps and velocity maps
Maps of integrated line intensity for each of the lines with extended emission are shown in Figs.~\ref{fig:mapsCO_N_HCOplus}--\ref{fig:mapsCH3OH_C2H}. Intensity is integrated over the velocity interval indicated in Table~\ref{t:maxint} for each map. The intervals are determined manually, with the aim of incorporating all emission belonging to the spectral line in question without including any signal from neighboring lines. Intervals are generally symmetric around the systemic \vlsr\ of $-5.8$\,\kms\ for AFGL2591, the interval width depending on the width of the line (CO~3--2 integrated over $[-29, 6]$\,\kms\ is the most extreme case). In the maps of integrated line intensity (Figs.~\ref{fig:mapsCO_N_HCOplus}--\ref{fig:mapsCH3OH_C2H}), contours are drawn based on resampled maps with pixels of 3\farcs75. Contour levels are scaled to the maximum intensity in each map listed in Table~\ref{t:maxint}. For comparison, SCUBA 850\,\micron\ continuum emission is shown on the background of each integrated line intensity map. In addition, we show an example of a map which is not extended according to our criterion: \SOtwo\ $15_{2,14}$--$14_{1,13}$ in Fig.~\ref{fig:notextended}. In any further discussion of \SOtwo\ morphology, we do not consider the $15_{2,14}$--$14_{1,13}$ mentioned here, but only the three transitions $5_{3,3}$--$4_{2,2}$, $23_{3,21}$--$23_{2,22}$, and $23_{2,22}$--$23_{1,23}$ that are listed in Table~\ref{t:spatialextent}.

\begin{table*}
\caption{Molecular lines with observed spatially extended emission. }             
\label{t:spatialextent}   
%\centering                                      
\begin{tabular}{l l l r@{ $\pm$ } l r@{ $\pm$ } l r@{ $\pm$ } l r@{ $\pm$ } l r@{ $\pm$ } l}            
\hline\hline
% header:
Molecule	& Transition	& Frequency\tablefootmark{a}	& \multicolumn{2}{c}{$\Delta$RA\tablefootmark{b}}	& \multicolumn{2}{c}{$\Delta$Dec\tablefootmark{b}}	& \multicolumn{4}{c}{Angular extent (Gaussian FWHM)\tablefootmark{c}}	& \multicolumn{2}{c}{Position angle\tablefootmark{d}}\\
\cline{8-11}  % underline "angular extent (Gaussian FWHM)" column header
			& 			& (MHz)	&  \multicolumn{2}{c}{(\arcsec)} & \multicolumn{2}{c}{(\arcsec)} &	\multicolumn{2}{c}{Major axis ($\arcsec$)} & \multicolumn{2}{c}{Minor axis ($\arcsec$)} & \multicolumn{2}{c}{(degrees)} \\

\hline

% table data (without FREQ and TeX labels) from extenttabletotex.py
          CO &          $J = 3-2$ & 345796.0
           &   $1.0$ &   $0.2$ &   4.7 &   0.4 &  84.6 &   1.6 &  53.9 &   0.6 &   94.65 &  0.02 \\
        \thCO &          $J = 3-2$ & 330588.0	
         &  $2.7$ &   0.2 &  $-1.4$ &   0.2 &  53.9 &   0.6 &  46.6 &   0.5 &  121.88 &  0.05 \\
        \CstO &         $J =  3-2$ & 337061.2\tablefootmark{e}	
         &  $-2.9$ &   0.2 &  $-2.6$ &   0.2 &  42.4 &   0.6 &  30.5 &   0.4 &   78.05 &  0.03 \\
 \CCH &       $N_J = 4_{9/2}-3_{7/2}$ & 349338.3\tablefootmark{f}
 	 &  $-2.5$ &   0.2 &  $-1.3$ &   0.1 &  30.7 &   0.5 &  22.0 &   0.3 &  105.91 &  0.03 \\
\CCH &       $N_J = 4_{7/2}-3_{5/2}$ & 349340.0\tablefootmark{g}	
	 &  $-2.8$ &   0.2 &  $-0.9$ &   0.1 &  29.1 &   0.4 &  20.6 &   0.3 &  108.54 &  0.03 \\
\methanol & $J_K = 7_1-6_1\ A^+$ & 	335582.0
	 &   $0.2$ &   0.4 &  $-3.4$ &   0.4 &  23.8 &   1.1 &  13.7 &   0.6 &  135.41 &  0.06 \\
\methanol & $J_K = 12_1-12_0\ A^\mp$   & 336865.1
	 &  $-1.7$ &   0.7 & $-1.7$ &   0.6 &  21.2 &   1.8 &  13.1 &   1.1 &  118.79 &  0.12 \\
\methanol & $J_K = 7_0-6_0\ E$     & 338124.5
	 &   1.3 &   0.3 &  $-2.6$ &   0.3 &  21.5 &   0.8 &  17.5 &   0.7 &   34.25 &  0.13 \\
\methanol & $J_K = 7_{-1} $-$ 6_{-1}\ E$ & 338344.6 
	 &   3.2 &   0.4 &  $-2.2$ &   0.3 &  30.3 &   1.0 &  20.6 &   0.7 &   89.08 &  0.06 \\
\methanol & $J_K = 7_0-6_0\ A^+$ & 338408.7
	 &   4.7 &   0.4 &  $-2.3$ &   0.3 &  31.2 &   0.9 &  23.0 &   0.7 &   70.66 &  0.07 \\
\methanol & $J_K = 7_2-6_2\ E$\tablefootmark{h} & 338722.3\tablefootmark{h} 
	 &   0.7 &   0.3 & $-3.2$ &   0.3 &  19.6 &   0.8 &  15.5 &   0.6 &  114.61 &  0.12 \\
\methanol & $J_K = 4_{0}-3_{-1}\ E$ & 350687.7
	 &   1.8 &   0.3 &  $-1.4$ &   0.3 &  34.6 &   0.8 &  25.3 &   0.6 &   72.59 &  0.05 \\
\methanol & $J_K = 1_{1}-0_{0}\ A^+$ & 350905.1
	 &   2.3 &   0.3 &  $-0.6$ &   0.3 &  33.1 &   0.8 &  26.5 &   0.6 &   75.68 &  0.07 \\
\methanol & $J_K = 4_1-3_0\ E$ & 358605.8 
	 &   1.0 &   0.3 &  $-2.1$ &   0.2 &  24.5 &   0.8 &  15.9 &   0.5 &    0.36 &  0.05 \\
\methanol & $J_K = 7_2-6_1\ E$ & 363739.8
	 &   1.7 &   0.3 &  $-0.4$ &   0.3 &  19.0 &   0.8 &  15.1 &   0.6 &  143.75 &  0.13 \\
 CN & $N_J = 3_{7/2}-2_{5/2}\ \Delta F$=0 & 340263.7\tablefootmark{i}
	 &  $-1.3$ &   0.4 &  $-5.4$ &   0.3 &  22.3 &   1.1 &  17.3 &   0.8 &    3.03 &  0.13 \\
CN & $N_J = 3_{5/2}-2_{3/2}\ \Delta F$=1	& 340033.5\tablefootmark{j}
	 &  $-1.4$ &   0.1 &  $-4.8$ &   0.1 &  24.0 &   0.3 &  19.4 &   0.2 &  105.12 &  0.04 \\
CN & $N_J = 3_{7/2}-2_{5/2}\ \Delta F$=1 & 340248.2\tablefootmark{k}
	 &   0.2 &   0.2 &  $-5.7$ &   0.1 &  27.5 &   0.4 &  20.7 &   0.3 &   93.23 &  0.03 \\
 CS &          $J = 7-6$ & 	342882.9 
	 &   0.1 &   0.1 &   1.3 &   0.1 &  31.3 &   0.4 &  21.1 &   0.2 &   82.87 &  0.02 \\
\CtfS &    $J = 7-6$ & 337396.5	
	 &  $-4.0$ &   0.4 &   0.7 &   0.3 &  20.9 &   1.0 &  13.0 &   0.6 &  117.48 &  0.07 \\
\mbox{o-\HHCO} &  $J_{K_a,K_c} = 5_{1,5}-4_{1,4}$  & 351768.6
	 &  $-0.8$ &   0.2 &  $-2.7$ &   0.1 &  28.1 &   0.4 &  19.9 &   0.3 &  179.56 &  0.03 \\
\mbox{p-\HHCO} &   $J_{K_a,K_c} = 5_{0,5}-4_{0,4}$ & 362736.0
	 &  $-0.6$ &   0.2 &  $-2.7$ &   0.1 &  22.8 &   0.4 &  16.1 &   0.3 &  169.49 &  0.04 \\
\mbox{p-\HHCO} &  $J_{K_a,K_c} = 5_{2,3}-4_{2,2}$ & 365363.4
	 &   1.4 &   0.2 &  $-1.5$ &   0.2 &  21.1 &   0.5 &  14.0 &   0.3 &   20.99 &  0.04 \\
      HCN &          $J = 4-3$ & 354505.6\tablefootmark{l} 
	 &   1.9 &   0.1 &  $-0.9$ &   0.1 &  26.5 &   0.3 &  19.4 &   0.3 &   95.18 &  0.03 \\
H$^{13}$CN 	& $J = 4-3$ & 345339.7\tablefootmark{m}
	 &   2.4 &   0.1 &  $-3.3$ &   0.1 &  17.0 &   0.2 &  14.0 &   0.2 &  127.44 &  0.05 \\
HCO$^+$ &          $J = 4-3$ &356734.2
	 &  $-1.2$ &   0.1 &  $-1.5$ &   0.1 &  31.3 &   0.3 &  22.2 &   0.2 &   75.62 &  0.02 \\
H$^{13}$CO$^+$ &    $J = 4-3$ & 346998.3
	 &   0.6 &   0.1 &  $-0.8$ &   0.1 &  23.7 &   0.3 &  18.2 &   0.3 &   93.28 &  0.04 \\
        HNC &          $J = 4-3$ & 362630.3
	 &  $-2.1$ &   0.1 &  $-4.1$ &   0.1 &  25.8 &   0.2 &  17.4 &   0.1 &   75.90 &  0.01 \\
\NtwoHplus &          $J = 4-3$ & 372672.5
	 &  $-3.4$ &   0.3 &  $-3.1$ &   0.2 &  40.1 &   0.9 &  17.7 &   0.4 &   69.61 &  0.02 \\
 \SOtwo & $J_{K_a,K_c} = 23_{3,21}-23_{2,22}$ & 336089.2
	 &  2.1 &  0.4 & $-2.1$ &  0.5 &  18.8 &   1.2 &  12.8 &   0.8 &   55.95\tablefootmark{n} &  0.11 \\
  \SOtwo & $J_{K_a,K_c} = 5_{3,3}-4_{2,2}$ & 351257.2
	 & $-0.3$ &  0.1 & $-3.1$ &  0.1 &  18.0 &   0.2 &  13.8 &   0.2 &  141.41 &  0.04 \\
   \SOtwo & $J_{K_a,K_c} = 23_{2,22}-23_{1,23}$ & 363925.8
	 &  0.8 &  0.4 &  0.9 &  0.5 &  21.4 &   1.3 &  10.5 &   0.6 &   58.52\tablefootmark{n} &  0.05 \\
       SO &       $N_J = 7_8-6_7$ & 340714.2
	 &   3.1 &   0.1 &  $-5.3$ &   0.1 &  17.8 &   0.3 &  15.2 &   0.3 &  118.64 &  0.08 \\
          SO &       $N_J = 8_8-7_7$ & 344310.6
	 &   2.0 &   0.1 &  $-2.8$ &   0.1 &  18.7 &   0.2 &  16.9 &   0.2 &  152.92 &  0.07 \\
         SO &       $N_J = 9_8-8_7$ & 346528.5
	 &   1.3 &   0.1 & $-1.7$ &   0.1 &  22.0 &   0.3 &  17.5 &   0.2 &  109.09 &  0.03 \\
\hline
continuum	& $850$\,\micron		& (N/A)	& $-1.2$ & $0.1$ & $-1.0$ & $0.1$ & $36.6$ & $0.2$ & $27.4$ & $0.2$ & $65.51$ & $0.01$ \\
\hline

\end{tabular}
\tablefoot{Spatial extent determined from two-dimensional Gaussian fit to integrated intensity map of each transition. Uncertainties for peak position, major and minor axes, and position angle quoted here are formal fitting errors and do not include systematic errors due to pointing uncertainties.  }
\tablefoottext{a}{Laboratory frequency of line transition from CDMS \citep{muller2005}.} \\
\tablefoottext{b}{Offsets in right ascension ($\Delta$RA) and declination ($\Delta$Dec) with respect to the reference position 20$^\mathrm{h}$29$^\mathrm{m}$24\fs9, 40\degr11\arcmin19\arcsec\ (J2000). } \\
\tablefoottext{c}{No beam deconvolution is applied before determining angular sizes.} \\
\tablefoottext{d}{Position angle of the fitted major axis, expressed from north to east.} \\
\tablefoottext{e}{Blend of fourteen hyperfine components (total separation 1.6\,MHz).} \\
\tablefoottext{f}{Blend of $F$=5--4 and 4--3 hyperfine components (separation 1.3\,MHz).} \\
\tablefoottext{g}{Blend of $F$=4--3 and 3--2 hyperfine components (separation 1.4\,MHz).} \\
\tablefoottext{h}{Blended with $7_{-2}-6_{-2}$ transition (separation 1.3\,MHz). The frequency listed in the table is the average.} \\
\tablefoottext{i}{Blended with OS$^{18}$O. Listed frequency is the average for the OS$^{18}$O $5_{3,3}$--$4_{2,2}$ transition and the $F$=5/2--5/2 and 7/2--7/2 hyperfine components of CN $3_{7/2}-2_{5/2}$ (total separation 3.2\,MHz).} \\
\tablefoottext{j}{Blend of $F$=7/2--5/2, 3/2--1/2 and 5/2--3/2 hyperfine components (total separation 4\,MHz).} \\
\tablefoottext{k}{Blend of $F$=7/2--5/2, 9/2--7/2 and 5/2--3/2 hyperfine components (total separation 0.8\,MHz).} \\
\tablefoottext{l}{Blend of $F$=4--4, 3--2, 4--3, 5--4, 3--4 and 3--3 hyperfine components (total separation 3.6\,MHz).} \\
\tablefoottext{m}{Blend of \SOtwo\ $13_{2,12}-12_{1,11}$ and hyperfine components $F$=4--4, 3--2, 4--3, 5--4, 3--4 and $F$=3--3 of H$^{13}$CN 4--3 (total separation 3.6\,MHz).} \\
\tablefoottext{n}{Position angle poorly represent actual morphology of the map (see Fig.~\ref{fig:mapsH2CO_SOx_CS}).}
\end{table*}

\begin{table*}
\caption{Maximum integrated intensity values and velocity parameters.  }             
\label{t:maxint}   
%\centering                                      
\begin{tabular}{l l r l r r@{ $\pm$ } l r@{ $\pm$ } l}   
\hline
Molecule	& Transition		&  $E_\mathrm{up}/k$	& integration                         & maximum 	& \multicolumn{4}{c}{Gaussian line profile}  \\
\cline{6-9}
		&				&					& interval\tablefootmark{a} & \intintens \tablefootmark{b} & \multicolumn{2}{c}{ $V_\mathrm{centroid}$\tablefootmark{c} }	& \multicolumn{2}{c}{FWHM\tablefootmark{d} }	 \\
		&				& (K)					& (\kms)				& (\Kkms)     	& \multicolumn{2}{c}{ (\kms) }				& \multicolumn{2}{c}{ (\kms) }    \\
\hline\hline
										                                                                                                                                      % (gaussian surface)
          CO &          $J = 3-2$ 			&33.2		& $[-29, +6]$		& 410.9		& $-4.36$	& $0.29$	& $4.8$	& $0.6$\tablefootmark{e} \\ % 226 = 55%
        \thCO &          $J = 3-2$ 			& 31.7		& $[-18, +6]$		& 225.5		& $-5.92$	& $0.11$	& $5.4$	& $0.3$\tablefootmark{e}	\\ % 173 = 77%
        \CstO &         $J =  3-2$ 			& 32.4		& $[-16, +4]$		& 18.6		& $-5.86$ & $0.06$	& $3.6$	& $0.2$\tablefootmark{e}	 \\ % 16.1 = 87%
 \CCH & $N_J = 4_{9/2}-3_{7/2}$		& 41.9 		& $[-20, +6]$		& 17.2		& $-5.99$	& $0.09$	& $4.3$	& $0.3$\tablefootmark{e}	\\ %13.0 = 76%
\CCH &  $N_J = 4_{7/2}-3_{5/2}$		& 41.9 		& $[-15, +4]$ 		& 14.0		& $-5.81$	& $0.13$	& $4.8$	& $0.3$\tablefootmark{e} \\ % 10.6 = 76%
\methanol & $J_K = 7_1-6_1\ A^+$		  & 79.0	             & $[-12, $-$1]$             & 2.7             & $-5.92$ & $0.12$  & $4.0$  & $0.3$ \\ % 2.2 = 81%
\methanol & $J_K = 12_1-12_0\ A^\mp$  & 197.1	      & $[-11, 0]$              & 1.7                 & $-5.36$ & $0.19$  & $3.5$  & $0.5$ \\ % 1.45 = 85%
\methanol & $J_K = 7_0-6_0\ E$              & 78.1	             & $[-10, $-$1]$         & 2.9                 & $-5.60$ & $0.11$  & $3.5$  & $0.3$ \\ % 2.1 = 72%
\methanol & $J_K = 7_{-1} $-$ 6_{-1}\ E$ 	  & 70.5	             & $[-11, 0]$       & 3.3                 & $-5.93$  & $0.05$  & $3.3$  & $0.2$ \\ % 3.4 = 103%
\methanol & $J_K = 7_0-6_0\ A^+$ 	  & 65.0		      & $[-14, +1]$            & 4.1                 & $-5.80$  & $0.07$  & $3.7$  & $0.2$ \\  % 3.5 = 85%
\methanol & $J_K = 7_2-6_2\ E$             & 87.3 		      & $[-11, +1]$            & 3.1                 & $-5.60$  & $0.07$  & $3.8$  & $0.2$ \\ % 2.6 = 84%
\methanol & $J_K = 4_{0}-3_{-1}\ E$       & 36.3               & $[-16, +1]$            & 5.8                 & $-7.01$  & $0.13$  & $5.7$  & $0.3$ \\ % 4.8 = 83%
\methanol & $J_K = 1_{1}-0_{0}\ A^+$ 	  & 16.8	             & $[-11, 0]$              & 3.5                 & $-6.05$  & $0.05$  & $3.5$  & $0.2$ \\ % 3.0 = 86%
\methanol & $J_K = 4_1-3_0\ E$ 	         & 44.3	             & $[-11, 0]$              & 3.5                 & $-6.42$  & $0.07$  & $3.8$  & $0.2$ \\ % 2.8 = 80%
\methanol & $J_K = 7_2-6_1\ E$ 		  & 87.3	             & $[-10, 0]$              & 2.6                 & $-5.55$  & $0.09$  & $3.2$  & $0.3$ \\  % 2.2 = 85%
CN & $N_J = 3_{7/2}-2_{5/2}\ \Delta F$=0 & 32.7		& $[-13, $-$1]$		& 3.3                 & $-5.38$  & $0.24$  & $6.6$  & $0.6$ \\  % 3.5 = 106%
CN & $N_J = 3_{5/2}-2_{3/2}\ \Delta F$=1 & 32.6		& $[-16, 0]$		& 21.5     		& $-5.80$  & $0.19$  & $6.6$   & $0.5$ \\ % 17.7 = 82%
CN & $N_J = 3_{7/2}-2_{5/2}\ \Delta F$=1 & 32.7		& $[-14, +2]$		& 25.7               & $-5.61$  & $0.08$  & $4.0$  & $0.2$ \\  % 20.2 = 79%
CS &          $J = 7-6$ 				& 65.8		& $[-14, +3]$ 		& 27.3		& $-5.91$	& $0.04$	& $3.5$	& $0.1$\tablefootmark{e} 	 \\ % 19.5 = 71%
\CtfS &    $J = 7-6$  		    			   & 50.2           & $[-12, $-$1]$             & 2.6 		& $-5.68$	& $0.07$	& $3.3$	& $0.2$ \\  % 2.3 = 88%
o-\HHCO &  $J_{K_a,K_c} = 5_{1,5}-4_{1,4}$  	& 62.5	& $[-13, +2]$		& 17.9 		& $-5.91$	& $0.07$	& $3.9$	& $0.2$ \\  % 15.0 = 84%
p-\HHCO &   $J_{K_a,K_c} = 5_{0,5}-4_{0,4}$ 	& 52.4	& $[-11, 0]$		& 8.3		& $-5.62$	& $0.05$	& $3.4$  & $0.1$ \\  % 7.1 = 86%
p-\HHCO &  $J_{K_a,K_c} = 5_{2,3}-4_{2,2}$ 	& 99.7	& $[-13, +2]$		& 3.5		& $-5.48$	& $0.09$	& $3.8$	& $0.3$ \\  % 2.8 = 80%
HCN &          $J = 4-3$ 				& 42.5		& $[-17, +5]$		& 67.5		& $-5.68$	& $0.06$	& $5.3$	& $0.2$\tablefootmark{e} \\  % 55.9 = 83%
H$^{13}$CN 	& $J = 4-3$ 			& 41.4		& $[-15, +3]$		& 10.5		& $-5.32$	& $0.07$	& $5.8$	& $0.2$ \\  % 8.61 = 82%
HCO$^+$ &          $J = 4-3$ 			& 42.8		& $[-18, +5]$		& 100.7 		& $-6.10$	& $0.08$	& $5.2$	& $0.2$\tablefootmark{e} \\ % 94 = 93%
H$^{13}$CO$^+$ &    $J = 4-3$		& 41.6		& $[-13, 0]$		& 7.9		& $-6.03$	& $0.07$	& $3.1$	& $0.2$ \\ % 6.1 = 77%
HNC &          $J = 4-3$ 				& 43.5		& $[-12, +1]$		& 21.4		& $-5.67$	& $0.05$	& $3.4$	& $0.1$ \\ % 17.6 = 82%
\NtwoHplus &          $J = 4-3$ 			& 44.7		& $[-11, $-$2]$		& 9.7		& $-5.66$ & $0.06$  & $3.1$  & $0.2$\tablefootmark{e} \\  % 5.9 = 61%
\SOtwo & $J_{K_a,K_c} = 23_{3,21}-23_{2,22}$ & 276.0	& $[-11, $-$1]$		& 2.6 		& $-5.64$	& $0.17$	& $3.2$	& $0.4$ \\  % 1.5 = 58%
\SOtwo & $J_{K_a,K_c} = 5_{3,3}-4_{2,2}$ 	& 35.9	& $[-11, 0]$		& 4.5		& $-5.65$	& $0.06$	& $5.3$	& $0.2$ \\ % 4.2 = 93% 
\SOtwo & $J_{K_a,K_c} = 23_{2,22}-23_{1,23}$ & 259.9	& $[-11, +1]$		& 2.2		& $-5.37$ & $0.17$	& $4.7$	& $0.4$ \\ % 1.9 = 86%
SO &       $N_J = 7_8-6_7$ 			& 81.2		& $[-13, +2]$		& 11.9		& $-6.01$	& $0.05$	& $5.2$	& $0.2$ \\ % 7.6 = 64%
SO &       $N_J = 8_8-7_7$ 			& 87.5		      & $[-14, +1]$            & 11.0              & $-6.08$ & $0.05$  & $5.2$   & $0.2$  \\ % 9.2 = 84%
SO &       $N_J = 9_8-8_7$ 			& 78.8 		& $[-15, +5]$ 		& 14.0		& $-5.91$	& $0.17$	& $5.8$	& $0.4$ \\ % 9.7 = 69%

\hline 
\end{tabular} \\
% tablenotes here
\tablefoottext{a}{Integration is done including the entire channel in which the boundary falls.} \\
\tablefoottext{b}{Uncertainties in absolute intensity calibration are 15\%. Maximum intensity levels are used to scale the contour levels in Figs.~\ref{fig:mapsCO_N_HCOplus}--\ref{fig:mapsCH3OH_C2H}.}\\
\tablefoottext{c}{Centroid velocity, with respect to \vlsr, determined from Gaussian fit to the velocity profile at the central spatial position. } \\
\tablefoottext{d}{Full width at half maximum determined from a Gaussian fit to the velocity profile at the central spatial position.} \\
\tablefoottext{e}{Apparent non-Gaussian line profile and/or presence of line wings outside of Gaussian.}
\end{table*}

% ------- description of spatial distribution ---------
\subsection{Spatial distribution of integrated line intensity}
\label{sec:results_distr}

% overall distribution
The emission of all molecular tracers from Table~\ref{t:spatialextent}, with the exception of CO, its isotopologs, and \SOtwo, is distributed over a central region $\sim$$20$--$50\arcsec$ in diameter (Figs.~\ref{fig:mapsCO_N_HCOplus}, \ref{fig:mapsH2CO_SOx_CS}, \ref{fig:mapsCH3OH_C2H}). The flattened morphologies of roughly half of the maps are consistent with a position angle of the major axis between $65\degr$ and $95\degr$ (Table~\ref{t:spatialextent}). 
% CO maps: two spatial scales
The only maps in which more extended emission is visible are those of the 3--2 transition of CO, \thCO\ and \CstO\ (Fig.~\ref{fig:mapsCO_N_HCOplus}), while \SOtwo\ emission is the most concentrated. Maps of CO and \thCO\ trace not only the smaller scale morphology already noted above for other molecules, but also an additional larger scale envelope (up to $\sim$$100\arcsec$) at lower intensity with a different orientation at a position angle of $\sim$20\degr. The large scale (0.07--0.30\,pc from the central position) extends to the region where previous modeling by \citet{vandertak2000jul} suggests the molecular hydrogen density to be $<$$10^5$\,\pccm\ and dust and gas temperatures $<$$40$\,K. 

% northern plume/warp
A `plume' extending to the north at position angle $\sim$10\degr\ is apparent in maps of CO, \CstO, HCN, HNC, \NtwoHplus, \HCOplus, \mbox{o-\HHCO}, and \methanol\ $1_1$--$0_0~A^+$, $7_0$--$6_0~A^+$ and $7_{-1}$--$6_{-1}~E$ (Figs.~\ref{fig:mapsCO_N_HCOplus}, \ref{fig:mapsH2CO_SOx_CS}, \ref{fig:mapsCH3OH_C2H}). Its intensity level is generally at $\sim$10\% of the maximum, but it is stronger in the \methanol\ lines relative to the peak intensity. The feature appears to emerge from a position roughly 15\arcsec\ east of the peak position. Alternatively, it might trace a warped structure of the collapsing molecular cloud. The plume is also apparent in the 850\,\micron\ continuum map (Fig.~\ref{fig:mapsCO_N_HCOplus}). We recognize the same feature in the \thCO\ 3--2 map in Fig.~3 of \citet{vandertak1999}. The northern warped plume is not present in maps of \CCH\ nor in any of the sulfur bearing molecules. This difference in observed morphology is not just an effect of signal-to-noise, since some of the latter transitions have peak line strengths (Table~\ref{t:maxint}) comparable to the maps that do show the plume. The velocity structure of this plume is destribed in Sect.~\ref{sec:results_velo}.

%%%
% map of a transition of SO2 that shows NO spatial extent
%%%

% two 'arms' (or one indent) on southern side
Maps of CO and isotopologs, CN~3$_{7/2}$--2$_{5/2}$\,$\Delta F$=1 (Fig.~\ref{fig:mapsCO_N_HCOplus}), CS, \mbox{o-\HHCO}\ (Fig.~\ref{fig:mapsH2CO_SOx_CS}), \CCH\ and \methanol\ 4$_{0,3}$-3$_{-1,3}$ (Fig.~\ref{fig:mapsCH3OH_C2H}) show a hint of two separate southern `arms' $\sim$30\arcsec\ south of their main peak, separated by $\gtrsim20\arcsec$ in the east-west direction. Alternatively, this feature can be described as one indentation.

% N2H+ distribution, four-arm structure
The spatial distribution of \NtwoHplus\ 4--3 emission (Fig.~\ref{fig:mapsCO_N_HCOplus}, top right panel) is strikingly different from those of other molecular transitions observed here. It is stretched significantly in the east-west direction, mostly due to the spatial separation of the velocity components described in Sect.~\ref{sec:results_velo}. Whereas various other species manifest two southern `arms' (see above), only \NtwoHplus\ shows the same on the northern side, giving rise to a four-directional morphology. 

% spatial peak offset w.r.t. dust and between molecules
% HNC and CN offset to southwest
The emission from HNC 4--3 and from the three CN transitions (Fig.~\ref{fig:mapsCO_N_HCOplus}) peaks at a position offset $\sim$3\arcsec\ to the southwest with respect to the dust emission and the other molecular maps. Most other molecular maps have peak positions at $\Delta$Dec between $-3$\arcsec\ and $-1$\arcsec, and the 850\,\micron\ dust emission peaks at $\Delta$Dec=$-1$\arcsec (Table~\ref{t:spatialextent}). 
% SO offset to southeast
Moreover, the peaks of the SO maps, especially $7_8$--$6_7$, appear to be shifted $\sim$$4$--$5$\arcsec\ to the southeast (Fig.~\ref{fig:mapsH2CO_SOx_CS}). 
% CS, C34S offset to the north
In contrast, the CS and \CtfS\ maps (Fig.~\ref{fig:mapsH2CO_SOx_CS}) have peak positions offset 2--3\arcsec\ to the north.
% C2H offset to northwest
Finally, the two \CCH\ maps, although not obvious from their fitted peak positions, are displaced to the northwest (Fig.~\ref{fig:mapsCH3OH_C2H}). 
% compare offsets to pointing errors
We note that these position offsets are only marginally significant compared to the pointing accuracy (Sect.~\ref{sec:obs}).

% double peak in warm CH3OH
A spatially resolved two-peak structure is apparent in the map of \methanol\ 12$_{1}$--12$_{0}$ ($E_\mathrm{up}$=197\,K) in the top right panel of Fig.~\ref{fig:mapsCH3OH_C2H}. The two peaks are separated by roughly 12\arcsec\ (12\,000\,AU). The southeast warm \methanol\ peak is associated to the infrared source detected by \citet[their Fig.~1]{vandertak1999}, the canonical center of the envelope of dust and gas. We discuss this two-peak structure further in Sect.~\ref{sec:nonmodeled}. The only other molecular transitions shown in this paper with an upper level energy above 100\,K are the two \SOtwo\ 23--23 transitions, where a double peak structure is not seen. The \NtwoHplus\ morphology does not coincide with the spatially separated warm methanol pockets, neither in angular separation, nor in position angle. 

% warm SO2 morphology perpendicular to outflow
The high-excitation $J$=23--23 transitions \SOtwo\ transitions in Fig.~\ref{fig:mapsH2CO_SOx_CS} appear to be stretched in a direction perpendicular to the outflow direction. The fitted position angle for these two maps does not reflect the actual orientation of the extended emission (cf.~Table~\ref{t:spatialextent} and Fig.~\ref{fig:mapsH2CO_SOx_CS}), perhaps due to the boxy non-Gaussian shape of the emission. The real position angle for these maps would plausibly be around 140\degr--150\degr.

%%%
% velocity maps
%%%%

\begin{figure*}
	\centering
	\resizebox{\hsize}{!}{\includegraphics{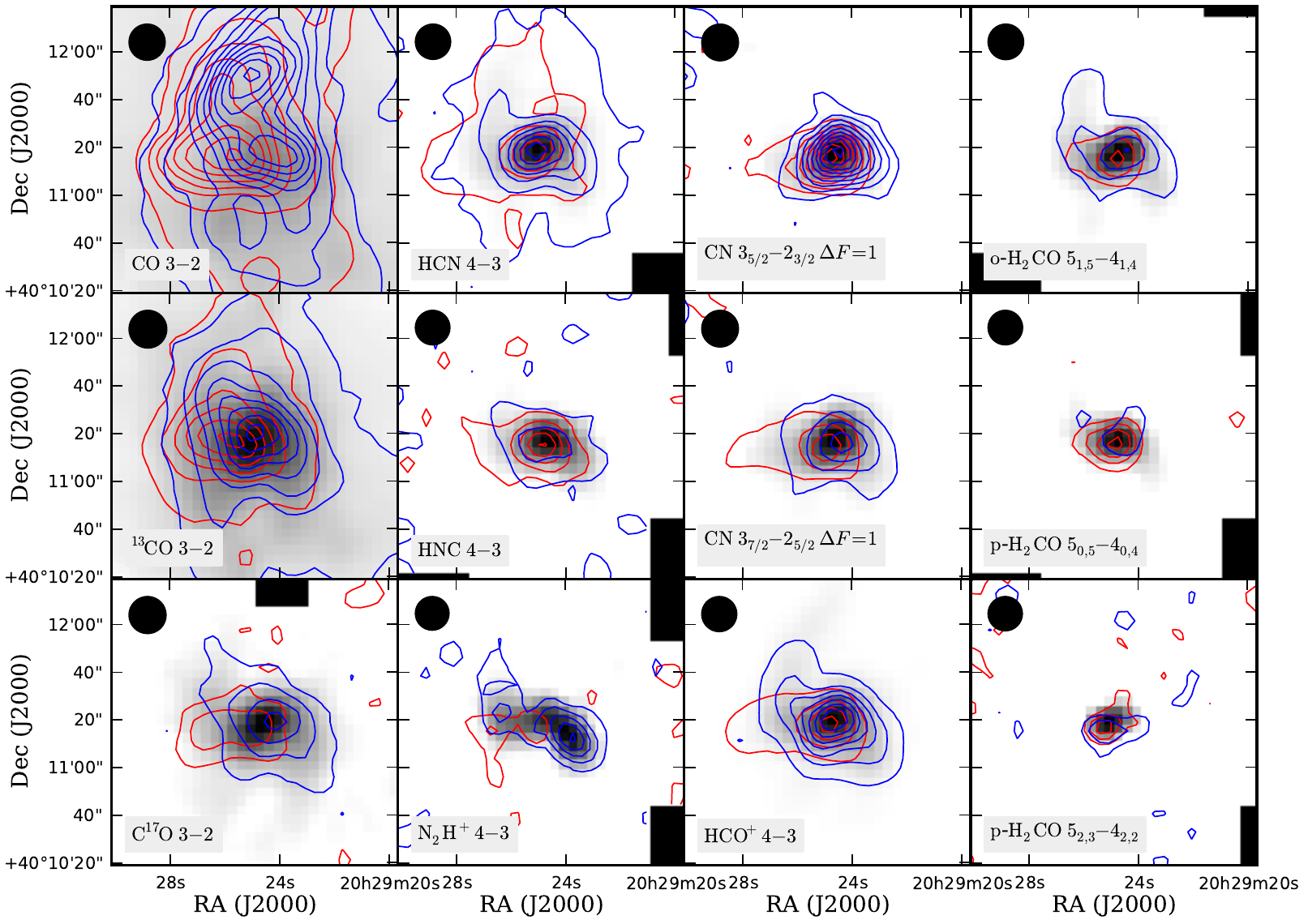}}
	\caption{Channel maps for the CO isotopologs, N-bearing species, \HCOplus\ and \HHCO. The velocity range is divided in three velocity bins with respect to the systemic velocity of $-5.8$\,\kms: $[-1.5, +1.5]$\,\kms\ is shown in grayscale stretched between 0.7\,\Kkms (white) and the maximum intensity in the central channel (black); all emission at relative velocities below $-1.5$\,\kms\ is shown in blue contours, and all emission at relative velocities above $+1.5$\,\kms\ in red contours. The lowest contour level is manually fine-tuned to emphasize velocity structure in each case without showing too much noise. Consecutive contour levels are drawn in steps of 10\% of the maximum intensity in the brightest channel. Lowest contours: 15.0\,\Kkms\ for CO, 5.0\,\Kkms\ for \thCO, 0.8\,\Kkms\ for \CstO, 0.35\,\Kkms\ for HCN and HNC, 0.4\,\Kkms\ for \NtwoHplus, 0.8\,\Kkms\ for CN $3_{5/2}$--$2_{3/2}$, 1.0\,\Kkms\ for CN $3_{7/2}$--$2_{5/2}$, and 0.8, 0.4 and 0.3\,\Kkms\ for the three respective \HHCO\ transitions. The FWHM of the telescope beam is indicated in the top left corner of each panel. 
	}
	\label{fig:velmapsCO_N_HxCO}
\end{figure*}

\begin{figure*}
	\centering
	\resizebox{\hsize}{!}{\includegraphics{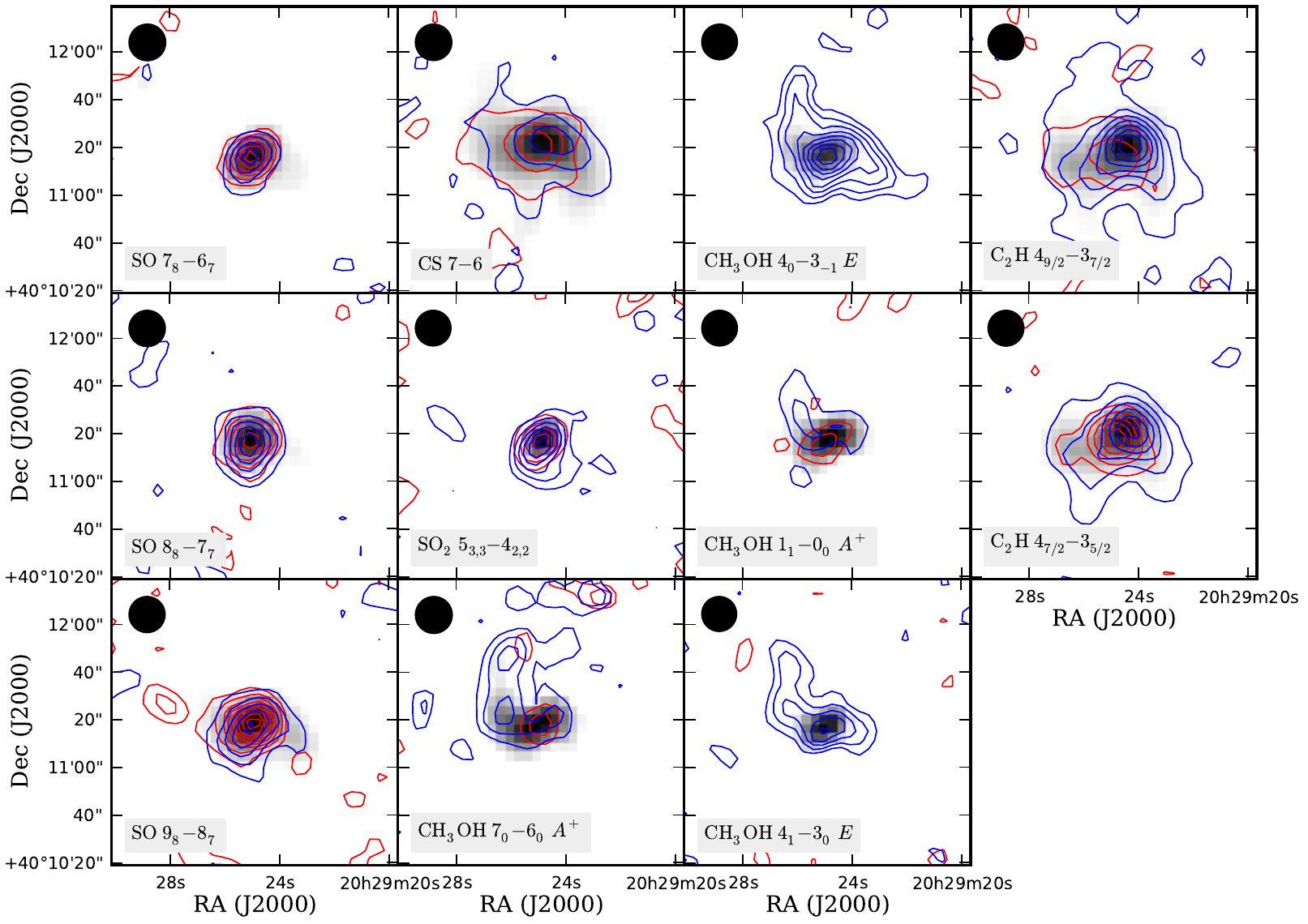}}
	\caption{Channel maps for sulfur bearing species, \methanol, and \CCH. Colors and contours have the same meaning as in Fig.~\ref{fig:velmapsCO_N_HxCO}, except lowest contours: 0.6\,\Kkms\ for SO $7_8$--$6_7$, 0.4\,\Kkms\ for SO $8_8$--$8_7$ and $9_8$--$8_7$, 0.6\,\Kkms\ for CS 7--6, 0.2\,\Kkms\ for \SOtwo\ $5_{3,3}$--$4_{2,2}$, 0.35, 0.6, 0.35 and 0.3\,\Kkms\ for the four consecutive \methanol\ transitions, and 0.4\,\Kkms\ for both \CCH\ transitions.}
	\label{fig:velmapsS_CH3OH_C2H}		
\end{figure*}

% ------ description of velocity structure -------
\subsection{Velocity structure}
\label{sec:results_velo}

% overall velocity profile of lines 
The velocity structure of the molecular gas can be illustrated by line profiles in velocity space, and by `channel maps'. First, the velocity profiles of all 35 lines at the central spatial position are fitted by a Gaussian to determine the centroid velocity and linewidth (Table~\ref{t:maxint}). We find centroid velocities between $-5$ and $-7$\,\kms, with a median of $-5.8$\,\kms. This is slightly lower than the previously determined velocity of $-5.5$\,\kms\ \citep{vandertak1999}. Gaussian linewidths range from $3$ to $7$\,\kms, with a median of 3.9\,\kms. 

\begin{figure*}
  \centering
  \resizebox{\hsize}{!}{\includegraphics{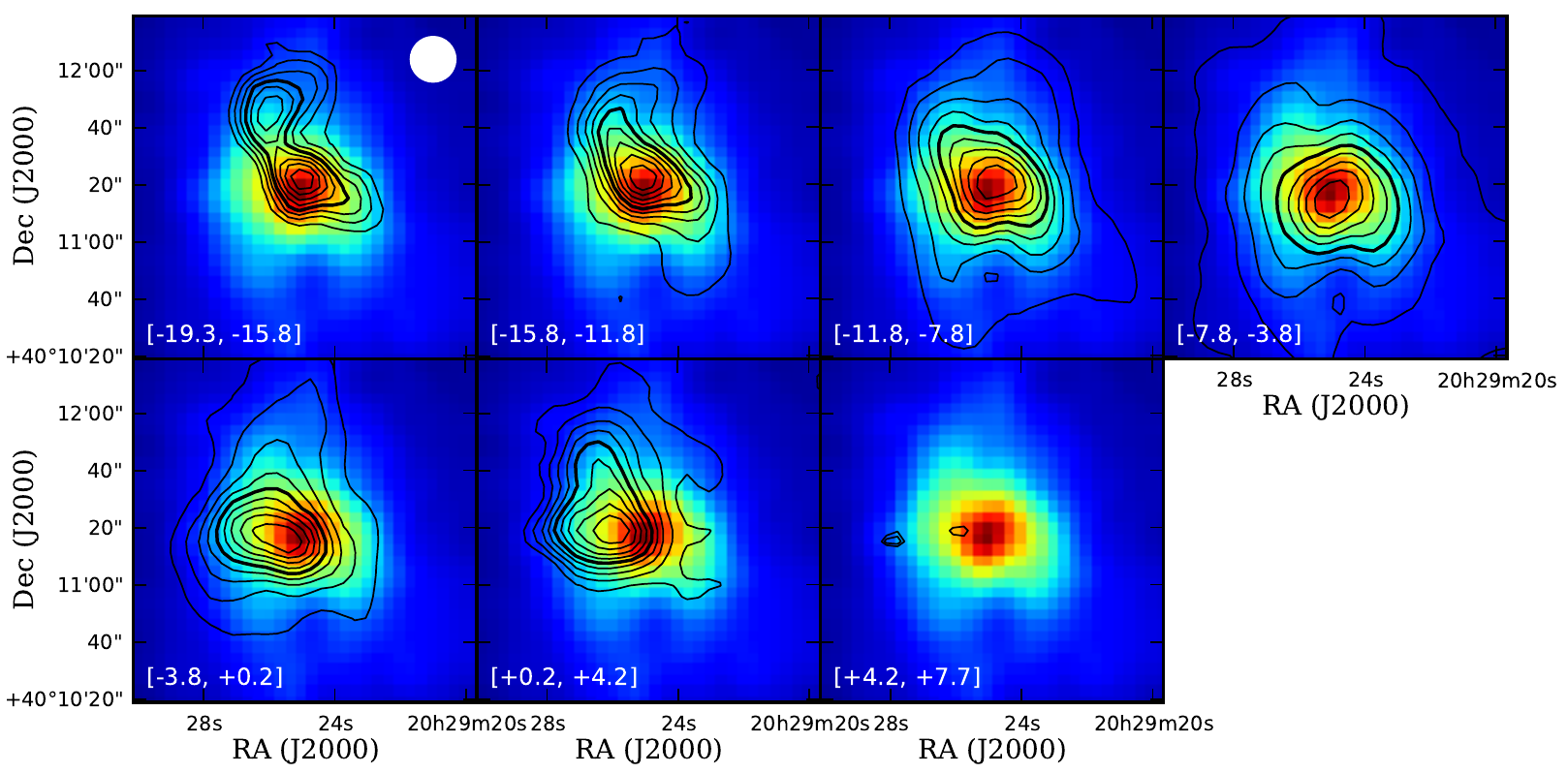}}
  \caption{Emission from the \thCO\ 3--2 transition divided over seven velocity bins. Line intensity is summed over the velocity range indicated in each panel (in \kms) and represented as black contour lines. Contour levels are 90\%, 80\%, \ldots, 10\% of the maximum integrated intensity in the velocity bin: 3.9, 14, 41, 124, 50, 8.1, and 0.95\,\Kkms\ in the respective bins. Like in Fig.~\ref{fig:mapsCO_N_HCOplus}, the 50\%-level contour is thicker, and any contours below 0.7\,\Kkms\ are not drawn. The total integrated \thCO~3--2 line intensity over the entire velocity range is shown in colorscale in each panel, for reference. }
  \label{fig:13COsevenbins}
\end{figure*}

% introduce 3-bin velocity maps
Second, to show the velocity structure of the overall molecular envelope, each integration interval from Table~\ref{t:maxint} is divided into three velocity bins: a central bin defined as the $[-1.5, +1.5]$\,\kms\ interval, relative to the systemic \vlsr\ of $-5.8$\,\kms, and `blue' and `red' bins composed of all emission below and above the central velocity bin, respectively. The resulting channel maps are presented in Figs.~\ref{fig:velmapsCO_N_HxCO} and \ref{fig:velmapsS_CH3OH_C2H}. We only show a map of velocity structure for those transitions that show significant emission in at least two of the three velocity bins. In addition, we present a velocity map where emission is divided over seven velocity bins for the bright \thCO~3--2 transition (Fig.~\ref{fig:13COsevenbins}).

% general velocity structure in 3-bin maps
A general west-east gradient from blueshifted (approaching) to redshifted (receding) emission is perceived in the CO species, \NtwoHplus, CN, \HCOplus, HCN, and HNC (Fig.~\ref{fig:velmapsCO_N_HxCO}). The same trend is seen in the red and blue peak positions in maps of \CCH\ and CS (Fig.~\ref{fig:velmapsS_CH3OH_C2H}). Although the typical angular separation between the peak of the blue and red components at $\lesssim$$10\arcsec$ is less than the FWHM of the telescope beam, the high signal-to-noise ratio in our maps allows for a relative position determination down to a fraction of the beam size. Moreover, the total extent of the blue and red lobes for the molecules mentioned above is significantly separated, with red contours reaching positions $\gtrsim 30\arcsec$ east, and blue contours $\gtrsim 30\arcsec$ west of the central position. Conversely, blue and red components are coincident for SO and \SOtwo. \HHCO\ maps show no clear velocity pattern. Velocity structure in maps of \methanol\ is complicated by blueshifted emission from the northeastern plume (see below, and Sect.~\ref{sec:results_distr}). Velocity structure in the spatially separated peaks of \methanol\ \mbox{12$_{1}$--12$_{0}$} is not discernable in our observations. Hence, the line of sight velocity of the two pockets must be within $1$\,\kms\ of each other.
% emission generally stronger on blue side
In addition, the majority of the velocity maps exhibits emission that is stronger in the blueshifted bin than in the redshifted bin. Exceptions are HNC~4--3 and the \mbox{p-\HHCO} transitions, where the red emission is stronger than the blue emission, and the SO transitions, where intensity levels in the red and blue bin are comparable.

% line widths and wings:
Upon inspection of the Gaussian line profiles parametrized by $V_\mathrm{centroid}$ and FWHM in Table~\ref{t:maxint}, the full integration interval is notably wider than the Gaussian profile for the CO species, as well as for \CCH, CS, HCN and \HCOplus. This is due to our original choice to incorporate all emission from a particular molecular transition in the corresponding \intintens\ map (Sect.~\ref{sec:selectextended}). The non-Gaussian part of the line intensity at the central spatial position mainly lies on the blueshifted side for these lines (see also Fig.~\ref{fig:HCOplus_vprofiles} and the notes in Table~\ref{t:maxint}).
For most of the lines mentioned above the line wings are relatively weak, comprising up to 20\% of the total integrated line intensity, but the single Gaussian profile is an especially poor representation for the main isotopic CO~3--2 line profile, where as much as 45\% of the integrated line intensity falls outside of the fitted Gaussian. We note that wings might be present even in other (weaker) lines, which would be invisible in the noise of the observations (Sect.~\ref{sec:obs}).

 % spatially resolved red/blue wings in N2H+
The velocity components of the \NtwoHplus~4--3 line are special in the sense that they are spatially resolved: the emission peaks of the blue and red \NtwoHplus wings in Fig.~\ref{fig:velmapsCO_N_HxCO} are separated by more than one beam size. For other species, the different components overlap in velocity space as well as spatially. Like for several other species, emission is significantly stronger in the blueshifted approaching component than in the receding component. In the four-arm structure noted for \NtwoHplus\ in Sect.~\ref{sec:results_distr}, the southeast arm consists exclusively of emission from the red velocity bin.

% come back to northern plume/warp
The northern plume noted in Sect.~\ref{sec:results_distr} has more contribution from the blueshifted component of the emission than from the redshifted and central velocity bins, as shown in Figs.~\ref{fig:velmapsCO_N_HxCO} and \ref{fig:velmapsS_CH3OH_C2H}. This is particularly clear in velocity maps of \methanol.

% ---- 13CO seven-bin channel map

% ---------- MODELING ---------
\section{Radiative transfer modeling}
\label{sec:modeling}

% explain new structure of this section
Section~\ref{sec:dustymodels} deals with dust continuum modeling and describes the derived density structure of the envelope. The spatial distribution of molecular emission from a radiative transfer model is addressed in Sect.~\ref{sec:static1Dmodels}. Section~\ref{sec:diffalpha} explores the effects of different density structures on molecular emission. Finally, the model from Sect.~\ref{sec:static1Dmodels} is adjusted to examine the effects of velocity structure (Sect.~\ref{sec:dynamicmodels}) and a flattened morphology (Sect.~\ref{sec:flatmodels}).

% ------- dust continuum models --------
\subsection{Dust continuum models}
\label{sec:dustymodels}

% summarize method of Jennifer's DUSTY models
We use the {\sc dusty} code \citep{ivezic1999} to create a grid of dust continuum models starting from the density profile of material in a spherical ($n \propto r^{-\alpha}$) envelope around a luminous source. Parameters that vary across the grid include the temperature at the inner edge of the dust envelope, and the power law index $\alpha$ of the density distribution. The internal luminosity is fixed at \pow{2}{4}\,$L_\odot$ at $1$\,kpc distance. A self-consistent dust temperature and radiation field is derived within the envelope, taking into account scattering, absorption and emission processes along a radial path. The intrinsically one-dimensional models are resampled onto a radial grid of sky coordinates to obtain an intensity map, which is subsequently convolved with a synthetic telescope beam. 

% summarize results of DUSTY models for AFGL2591
The resulting radial intensity profile of each model is compared with the 450 and 850\,\micron\ SCUBA continuum maps (Sect.~\ref{sec:obs}). The best-fit {\sc dusty} model for 450\,\micron\ has $\alpha=1.10$, whereas the one for 850\,\micron\ has $\alpha=1.45$. 
We take the modeled dust opacity, $\tau_\nu$, and use the relation
\begin{equation}
\tau_{\nu} = \int \kappa_{\nu} \rho(r) \mathrm{d}r
\end{equation}
to derive the scaling factor $\rho_0$ in the mass density, $\rho(r) = \rho_0 (r/r_0)^{-\alpha}$. Here, we assume values for the absorption coefficients $\kappa_{850} = 0.03\,\mathrm{cm}^2\,\mathrm{g}^{-1}$ and $\kappa_{450} = 0.07\,\mathrm{cm}^2\,\mathrm{g}^{-1}$. The mass density scaling factor $\rho_0$ is divided by the mass of a hydrogen molecule in order to convert to gas number density ($n_0$). The description of density and temperature used for the $\alpha=1.5$ models in Sects.~\ref{sec:diffalpha} and \ref{sec:dynamicmodels} is derived from an $\alpha=1.50$ {\sc dusty} model, which is very similar to the best-fit solution with $\alpha=1.45$. 

% justify alpha=1.0 density power-law
\citet{vandertak2000jul} constrained the slope of the power law density profile, $\alpha$, using CS line observations. With the index being allowed to take on values of $0.5$, $1.0$, $1.5$ or $2.0$, they found that the best fit to several observed CS lines is given by $\alpha=1.0$. The observed  450\,\micron\ continuum map also shows good correspondence to the radial intensity profile of an $\alpha=1.0$ model, but the 850\,\micron\ continuum map is fit better with $\alpha=1.5$ (Sect.~5.3 of \citealt{vandertak2000jul}). We have independently confirmed the result that 850\,\micron\ dust emission favors a steeper density gradient ($\alpha=1.45$) than 450\,\micron\ ($\alpha=1.10$). 
% mention density slope found by continuum modeling and by De Wit+ (2009) in mid-IR
In addition, \citet{dewit2009} find that a power law density profile with $\alpha=1.0$ fits the radial intensity profile of high-resolution ($0.3\arcsec$ half-power beamwidth) 24.5\,\micron\ observations, but they also report that a fit to the spectral energy distribution prefers $\alpha=1.5$. 
Based on the above arguments, we choose a density profile with $\alpha=1.0$ for the molecular radiative transfer in Sects.~\ref{sec:static1Dmodels}, \ref{sec:dynamicmodels}, \ref{sec:flatmodels}. An exploration of different values of $\alpha$ is addressed in Sect.~\ref{sec:diffalpha}.

% ---------- 1D-models -----------
\subsection{Static spherical model}
\label{sec:static1Dmodels}

% explain 1st order validity of spherical modeling and species selection 
In the quantitative modeling of the physical structure of the envelope that gives rise to the molecular emission, we focus on \CstO, HCN, CS, \CtfS, \HCOplus\ and \HthCOplus. This set of species is chosen to cover a variety of atomic constituents, and to have observed line strengths that provide the highest signal-to-noise available in our sample. While CS and \HCOplus\ are juxtaposed with rarer isotopic variants, the isotopolog H$^{13}$CN is not included because its line signal is blended with an \SOtwo\ transition (see Table~\ref{t:spatialextent}). The maps of \CstO, HCN, CS and \HCOplus\ have observed major to minor axis ratios below 1.5 (Table~\ref{t:spatialextent}), and they have a largely homogeneous spatial distribution. This makes them suitable tracers of the quiescent overall envelope of AFGL2591. 
% refer to where we discuss non-modeled species:
Section~\ref{sec:nonmodeled} addresses the interpretation of distributions for species that are not explicitly modeled here but do appear in Table~\ref{t:spatialextent}.

% adopted physical structure of the cloud
We use the radiative transfer code {\sc ratran} \citep{hogerheijde2000} to test one-dimensional source models for AFGL2591. This relies on molecular data files from {\sc lamda} \citep{schoeier2005}, with collisional rate coefficients calculated by \citet[HCN]{green1974}, \citet[CS]{turner1992}, \citet[\HCOplus]{flower1999}, and \citet[CO]{yang2010}. 
Motivated by the dust modeling described in Sect.~\ref{sec:dustymodels}, we adopt a physical model from \citet{vandertak1999,vandertak2000jul}, which assumes a power law density profile of the form:
\begin{equation}
n(r) = n_0 \left( \frac{r}{r_0} \right)^{-\alpha} , 
\label{eq:n1D}
\end{equation}
with $n_0$$=$\pow{5.3}{4}\,\pccm\ at $r_0$$=$\pow{2.7}{4}\,AU, and $\alpha$$=$$1$. The model cloud is truncated at a radius of \pow{6.2}{4}\,AU (62\arcsec), and has a total \HH\ mass of 193\,$M_\odot$. The innermost shell of our model extends from $r=0$ to $\sim$200\,AU, and has an \HH\ density of \pow{1.2}{7}\,\pccm\ and a temperature of 372\,K. The density singularity at $r=0$ is thus avoided by the discrete sampling over our finite number of model shells. A centrally peaked temperature profile ranging down to 25\,K was derived from dust emission by \citet{vandertak2000jul}. Molecular abundances are kept constant as a function of position in the envelope. A constant turbulence is incorporated, parametrized by the 1/$\mathrm{e}$ halfwidth velocity of 2\,\kms, consistent with the typical value of 3--4\,\kms\ of the measured FWHM linewidths (Table~\ref{t:maxint}). Initially, no infall, outflow, rotation or other velocity gradients are included. We refer to this model as the `static spherical model'. 

% production of maps
We use the ray tracing routine of {\sc ratran} to produce spectral maps with 0.5\arcsec\ pixels at the distance of 1\,kpc. These maps are intentionally oversampled with respect to the spatial resolution of our observations, and they are subsequently convolved with a Gaussian beam profile appropriate for the molecular transition under consideration. 

The spatial distribution of observed integrated line emission is probed by a slice along the `long axis' (position angle 75\degr) and another slice along the `short axis' (position angle 165\degr) of the flattened large scale cloud. These slices are compared with the static spherical model in Fig.~\ref{fig:static1Dmodels}. The two observed slices provide a measure of the non-spherical morphology of the cloud in the plane of the sky. Any modeled intensity profile which differs from the observed profiles by more than the difference between the observed `long axis' and `short axis' profiles should be regarded as incorrect.

% match peak intensities by varying abundance
We attempt to match the modeled peak line strengths to the observed ones by adjusting individual molecular abundance values, keeping isotopic ratios fixed at CS/\CtfS=20 and \HCOplus/\HthCOplus=60. We find abundances, relative to \HH, of \pow{2}{-9} for \CtfS\ and \pow{2.5}{-10} for \HthCOplus. Consequently, the abundances of CS and \HCOplus\ are then fixed at \pow{4}{-8} and \pow{1.5}{-8}, respectively. Since we do not have an unblended optically thin isotopolog for HCN, we take \pow{2}{-8}; for \CstO\ we adopt \pow{5}{-8} as an abundance. These abundance values are consistent with earlier estimates by \citet{vandertak1999}. 

% discuss match of thin lines, discrepancy for thick ones
The shapes of the modeled intensity profiles as a function of radius correspond to the observations as long as the line emission is predominantly optically thin (Fig.~\ref{fig:static1Dmodels}, \CtfS, \HthCOplus), but the emission of both HCN and \HCOplus\ grows optically thick before the modeled line intensity reaches the observed value. This results in modeled emission profiles which are too weak near the center of the cloud, and fall off less steeply than the observed intensity profile. A deviation from the fixed isotopic ratio is briefly investigated. However, since the cloud is already optically thick at the center, a higher molecular abundance for the main isotopic species does not increase the emerging line intensity at the central position, while the intensity at more optically thin lines of sight away from the center does increase, thereby \emph{adding} to the mismatch of the overall spatial profile. Moreover, with observed variations of the isotopic ratio restricted to a factor $\lesssim 3$ \citep[e.g.,][]{chin1996}, our results are not affected.

% overproduction of CS not significant
The modeled intensity of CS~7--6 is obtained by first adjusting the \CtfS\ abundance to match the observed peak intensity, and simply scaling the CS abundance by the canonical factor 20 (see above). The modeled main isotopic CS~7--6 emission appears to be stronger than observed, but we note that this discrepancy is well within the intensity calibration uncertainty of 15\% of the observations. 

\begin{figure}
  \centering
  \resizebox{\hsize}{!}{\includegraphics{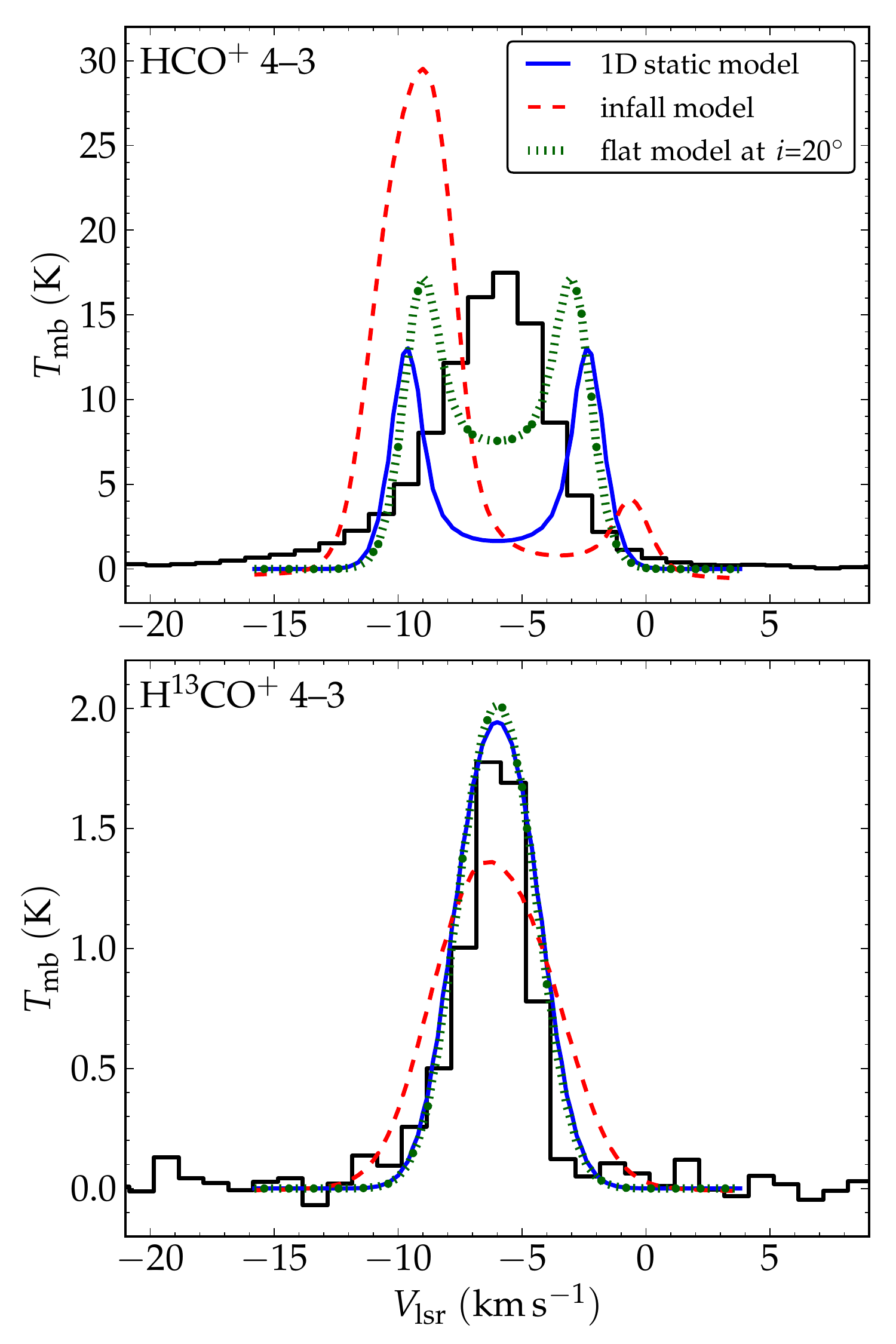}}
  \caption{\textit{Top:} Velocity profiles of observed \HCOplus~4--3 (histogram), compared to those from various models, all at the central spatial position. The profiles from the spherical (1D) model and the infall model follow from an \HCOplus\ abundance of \pow{1.5}{-8}, the abundance for the model used for the flattened model profile is \pow{6}{-9}. \textit{Bottom:} velocity profiles of observed and modeled \HthCOplus, with abundances scaled by a factor 60 with respect to H$^{12}$CO$^+$.  }
  \label{fig:HCOplus_vprofiles}
\end{figure}

\begin{figure}
  \centering
  \resizebox{\hsize}{!}{\includegraphics{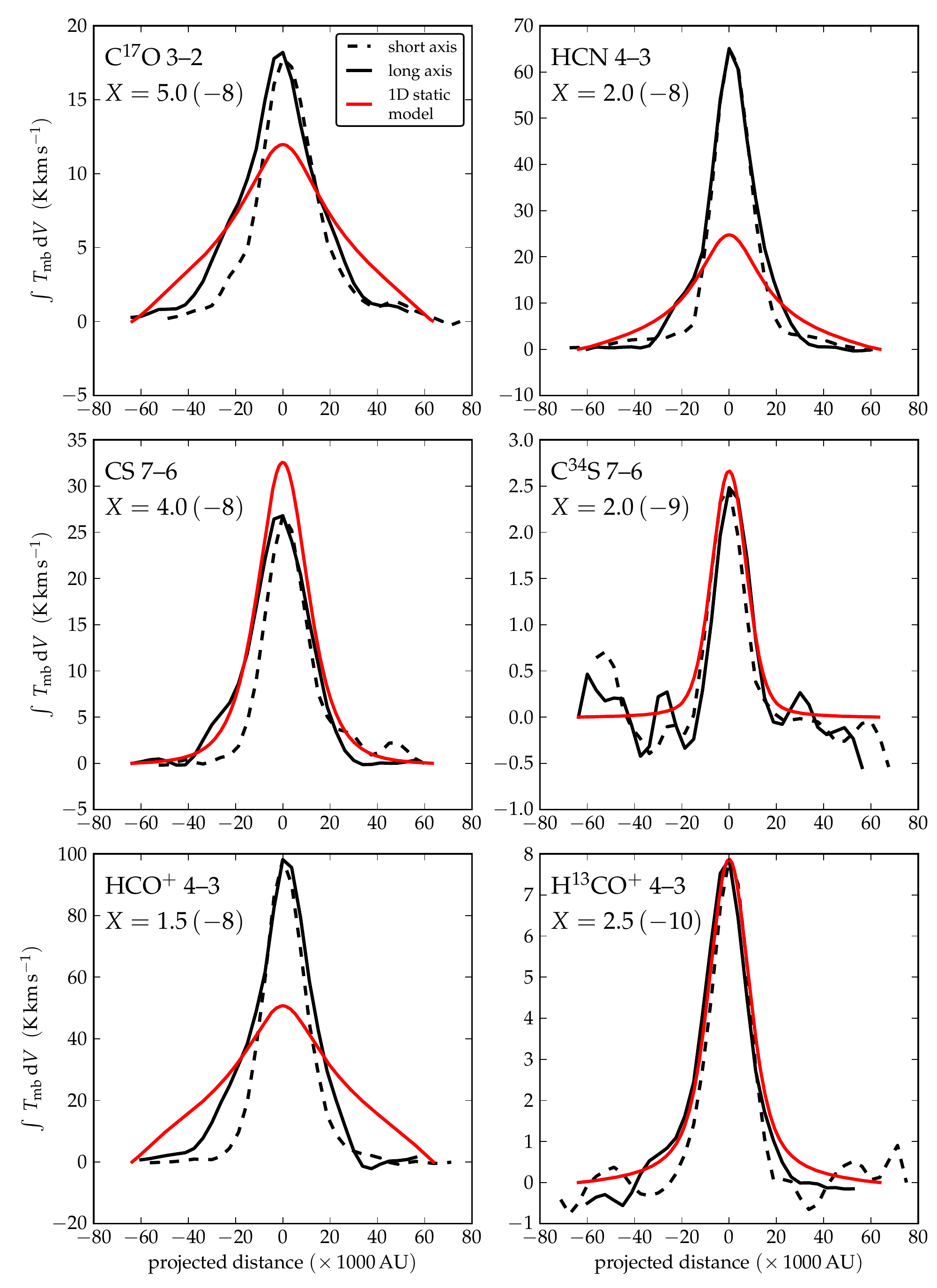}} 
  \caption{Position dependence of integrated line intensities for static spherical {\sc ratran} models (Sect.~\ref{sec:static1Dmodels}) of \CstO, HCN, CS, \CtfS, \HCOplus\ and \HthCOplus. Black solid and dashed lines represent two perpendicular slice directions in the observed maps, red lines represent radial dependence of integrated line strength in the 1D static models. At the line center, the {\sc sky} ray tracing routine calculates the following optical depths through the center of the model sphere: 0.095 for \CstO, 33 for HCN, 10 for CS, 0.38 for \CtfS, 34 for \HCOplus\ and 0.64 for \HthCOplus. Abundance values ($X$) are presented in the form $a\,(b)$, representing \pow{a}{b}. }
  \label{fig:static1Dmodels}
\end{figure}

% spectral line profiles match/mismatch
In the spectral dimension, each observed line has a roughly symmetric and single-peaked shape, as illustrated in Fig.~\ref{fig:HCOplus_vprofiles} for \HCOplus~4--3. Spectral line shapes are similar for other species. Considering the line profiles resulting from the radiative transfer models, individual optically thin lines (\CtfS~7--6, \HthCOplus~4--3, \CstO~3--2) have very comparable shapes, matching observed line shapes. However, for the main isotopologs HCN and \HCOplus, the high optical depth at line center is prominently displayed in the form of a self-absorbed line profile (Fig.~\ref{fig:HCOplus_vprofiles}).

% ---- panel of 1D static model profiles -----

% explain discrepant C17O, even though it is optically thin
A particular explanation is required for the spatial distribution of \CstO, which is also discrepant w.r.t.~the observed distribution but is optically thin in the modeled line. 
% rest of discussion moved to section 5
See Sect.~\ref{sec:discussmodels} for a discussion.

% propose velocity structure to diminish tau
To help the line emission escape the molecular cloud, the opacity at line center might be diminished for HCN and \HCOplus\ by introducing a velocity gradient in the protostellar cloud (Sect.~\ref{sec:dynamicmodels}), or by adjusting the geometry of the envelope (Sect.~\ref{sec:flatmodels}). In the originally optically thick cases this might result in extra emerging line intensity. On the other hand, the integrated line intensity of the optically thin \HthCOplus\ and \CtfS, for which the static spherical models already match the observations, should be conserved.

% ------- steeper density profiles --------
\subsection{Steeper density profiles}
\label{sec:diffalpha}

% remind that we chose alpha=1.0 for line ratran modeling
We recall that we adopt $\alpha=1$ in Sect.~\ref{sec:static1Dmodels} for the density profile given by Eq.~(\ref{eq:n1D}) for the static spherical model. The dynamic spherical models (Sect.~\ref{sec:dynamicmodels}) and the flattened models (Sect.~\ref{sec:flatmodels}) are based on the same density power law. The validity of this assumption is addressed in Sect.~\ref{sec:dustymodels}; below we illustrate how steeper density profiles affect our results. 

% discuss what happens with alpha=2.0 and alpha=1.5
Because the value of $\alpha$ is not unambiguously equal to $1.0$, we run test models with $\alpha=2.0$ and $\alpha=1.5$ for the static spherical model for \HCOplus. Both models have self-consistent temperature profiles derived in earlier work \citep{vandertak2000jul} or by the dust modeling described in Sect.~\ref{sec:dustymodels}. The qualitative trend is that a steeper density profile results in a steeper temperature gradient. 

% alpha=2 in spherical static model
In the static $\alpha=2.0$ model, the \HCOplus~4--3 integrated line intensity emanating from the center is about 20\% higher than that of the $\alpha=1.0$ model at the same abundance (cf.~Fig.~\ref{fig:static1Dmodels}). However, it still only reproduces 60\% of the observed peak integrated intensity. Doubling the \HCOplus\ abundance from \pow{1.5}{-8} to \pow{3}{-8} increases the central \intintens\ value by only a few percent, with the optical depth at the model center at line center increasing from 28 to 52. Therefore, even with further increased abundance, the $\alpha=2.0$ model will still underproduce the observed peak \intintens\ level. On the other hand, the spatial profile of \intintens\ does fall off steeper than for $\alpha=1.0$, bringing the shape closer to the observations, but we note that the \HthCOplus\ \intintens\ profile becomes too sharp to match the observations. Moreover, the fixed 1/60 ratio between the abundances of \HthCOplus\ and \HCOplus\ results in an \emph{over}production of the central \HthCOplus\ \intintens, simultaneously with the \emph{under}production of \HCOplus\ emission. 
Thus, we conclude that a density distribution with $\alpha=2.0$ does not improve the match to the observed intensity distribution with respect to the $\alpha=1.0$ case (Sect.~\ref{sec:static1Dmodels}). 
 
% alpha=1.5 model in spherical static case
In the static $\alpha=1.5$ model, the \HCOplus~4--3 integrated line intensity distribution is more peaked than in the $\alpha=1.0$ case from Sect.~\ref{sec:static1Dmodels}. However, the peak \intintens\ value, which was a factor $\sim$$2$ too low for $\alpha=1.0$, is increased by only $\lesssim$$10\%$ for $\alpha=1.5$. At the same time, \HthCOplus\ emission toward the central position is overproduced by a factor of $2$. We conclude, as for $\alpha=2.0$ and $1.0$, that a density power law slope of $1.5$ fails to explain the spatial distribution of line emission if we consider an optically thick and an optically thin tracer simultaneously.

% ------- dynamic models --------
\subsection{Dynamic spherical models}
\label{sec:dynamicmodels}

In this section we attempt to diminish the optical depth near the emission line frequencies by introducing velocity structure in the modeled cloud. Velocity structure can be considered as either random (turbulent) motion or systematic motion. We investigate the effects of a radial turbulence gradient and of a systematic infall signature on the line opacity and on the distribution of emerging intensity.

While the originally constant turbulence of 2\,\kms\ is consistent with non-thermal linewidths found in other massive star-forming regions \citep{caselli1995}, we introduce a gradient ranging from 10\,\kms\ in the central shell of the model down to 1\,\kms\ in the outer shell. Even with this deliberately extreme turbulence gradient, the resultant line optical depth is not reduced significantly, whereas the line shapes become broader in velocity space, as expected. Since this broadening is inconsistent with the observed FWHM of $\sim$4\,\kms\ for most lines, we refrain from further exploring this particular type of differential dynamics.

% introduce infalling cloud concept in context of literature
Apart from a turbulence gradient, a systematic radial velocity gradient could be viable. If our envelope is to be forming a centrally condensed object, an infalling rather than an expanding velocity field seems reasonable. 
% no double-peaked spectral line signatures in our data:
An infall signature is classically regarded as the double-peaked spectral line profile with a stronger blue than red peak in optically thick lines  (see, e.g.,~Fig.~5 of \citealt{evans1999}). Such a signature is either not present in our source, or is unresolved by the observations (spectral resolution 1\,\kms). Any unresolved double-peak structure would have to be separated by $<$2\,\kms, yielding an upper limit to the infall velocity of 1\,\kms\ near the dust sublimation radius, $\sim$100\,AU \citep[e.g.,][]{dewit2009}, close to the protostar. This limit is conservative, since even observations at slightly higher spectral resolution \citep{mitchell1992,vandertak1999} show no sign of infall in line profiles of, e.g., \thCO\ and CS.

% mention studies that have used spatial information to constrain infall models
\citet{choi1995} and \citet{hogerheijde2000ApJ} have considered one or more off-center positions to constrain models of infalling protostellar envelopes. They find that inside-out infall models with $\alpha=1.5$ are among the possible fits for some protostellar envelopes. We emphasize that different $\alpha$ values are found for different protostellar envelopes, although the power law index of the density profile $\alpha$ is generally between $1.0$ and $2.0$. \citet{hogerheijde2000ApJ} highlight that the density and velocity structure are degenerate.

% describe physical model for infalling case
To enable a direct comparison between results from the static and the infalling models, we retain the spherical $\alpha=1.0$ envelope from Sect.~\ref{sec:static1Dmodels} and add a radial velocity profile applicable for the freefall region of an inside-out collapsing cloud \citep[following][]{shu1977}:
\begin{equation}
V_\mathrm{infall}(r) = - \sqrt{\frac{2 G M(r)}{r}} ,
\label{eq:vshu}
\end{equation}
where $r$ represents radial distance from the cloud center and $G$ is the gravitational constant. The enclosed mass $M$ in Eq.~(\ref{eq:vshu}) is defined as $M(r) = M_\mathrm{central} + M_\mathrm{env}(<r)$. We set $M_\mathrm{central} = 16\,M_\odot$ to account for the central protostellar object; $M_\mathrm{env}(<r)$ represents the gas mass in the envelope contained in a sphere of radius $r$. The turbulent velocity is again constant at 2\,\kms\ in the infalling model. 
Relative molecular abundances are kept the same as in Sect.~\ref{sec:static1Dmodels}.

The velocity gradient has the desired effect of decreasing the optical depth at the line center and through the center of the spherical model by a factor $\gtrsim3$. 
As in the spherical static case (Sect.~\ref{sec:static1Dmodels}), the modeled profiles of the rarer isotopologs \CtfS\ and \HthCOplus\ still match the observations after the introduction of infall. The spatial profile of \intintens\ of HCN~$4$--$3$ and \HCOplus~$4$--$3$ benefit from the reduced optical depth. However, it is insufficient to reproduce the observed peak \intintens\ value for HCN, as well as the spatial profile of \HCOplus. On the other hand, the infall model for CS~7--6 overproduces line intensity at the central position by $\sim$$70\%$. In general, a significant discrepancy remains for the optically thick lines, both in terms of shape of the spatial profile as well as the peak integrated line intensity value. 
%
% unable to reconcile v-gradient with observed linewidth:
More importantly, the velocity gradient introduced in Eq.~(\ref{eq:vshu}) results in a very asymmetric and double-peaked velocity profile for optically thick lines, and a wider profile ($\mathrm{FWHM} \gtrsim 6$\,\kms, Fig.~\ref{fig:HCOplus_vprofiles}) for optically thin lines such as \HthCOplus~4--3. Concluding, the observed spectral line profiles indicate that an infall model is inadequate for the molecular envelope studied here.

\begin{figure}
  \centering
  \resizebox{\hsize}{!}{\includegraphics{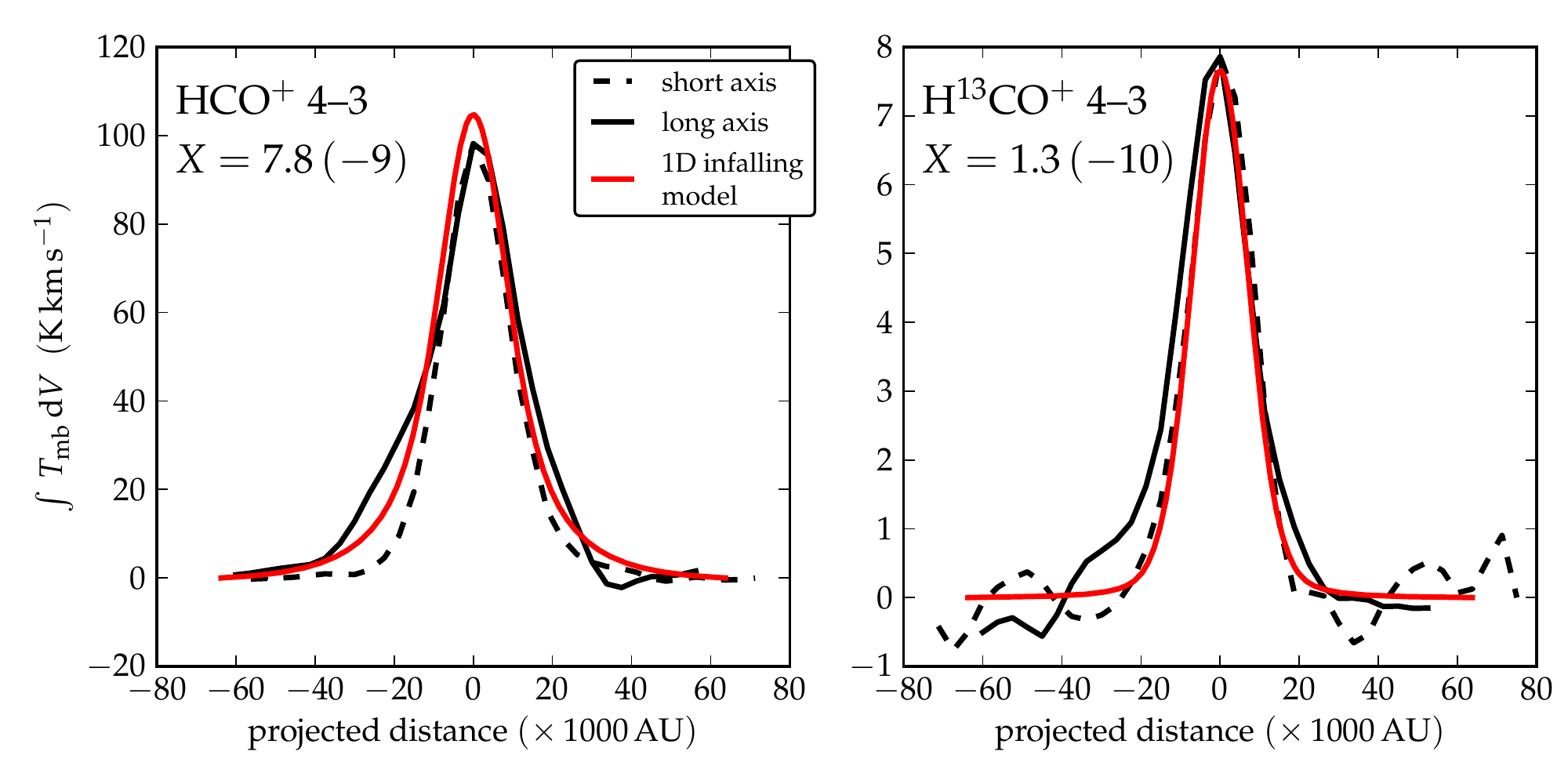}} 
  \caption{
  Position dependence of integrated line intensities for \HCOplus\ and \HthCOplus~4--3 resulting from an infall model with $\alpha=1.5$ (Sect.~\ref{sec:diffalpha}). Optical depths at line center through the center of the model cloud are $4.6$ for \HCOplus\ and $0.06$ for \HthCOplus. 
    }
  \label{fig:alpha15_infall}
\end{figure}

% alpha=1.5 model in spherical infalling case (moved from 'steeper density profiles' subsection)
Finally, we test an $\alpha=1.5$ model for the infalling case, dynamically consistent with the physical description by \citet{shu1977}. The spatial distribution of \intintens\ (Fig.~\ref{fig:alpha15_infall}) for this model comes closer than any other to a simultaneous explanation of an optically thick (\HCOplus) and an optically thin (\HthCOplus) tracer, with abundances halved with respect to the $\alpha=1.0$ models. However, when the spectral direction is also taken into account, the same discrepancies emerge as for the other infall models. We conclude that the adjustment of the density power law slope from $1.0$ to $1.5$ does not alleviate the mismatch between infall models and the observed spectral line profiles. In fact, the static models in Sects.~\ref{sec:static1Dmodels} and \ref{sec:flatmodels} produce better matches to the observed spectral line profiles than the infall models.

% ---------- 2D-models ---------
\subsection{Static model with flattened geometry}
\label{sec:flatmodels}

After considering a static spherical model (Sect.~\ref{sec:static1Dmodels}) and dynamic spherical models (Sect.~\ref{sec:dynamicmodels}), we now explore a morphological model of the molecular envelope which is flattened in one direction. The spatial axes of the model perpendicular to the line of sight are constrained by the observed morphology of the molecular cloud, which indicates a projected cloud structure not far from circular (major to minor axis ratios $<1.5$). On the other hand, the cloud can be flattened significantly in the direction along the line of sight. Moreover, if a small inclination of the short axis of the modeled system w.r.t.~the line of sight (`viewing angle') is introduced, we expect to be able to explain part of the non-circular geometry in the observed maps. Geometrically, the observed major/minor axis ratios on the sky of $\lesssim 1.5$ would constrain a flattened structure to be positioned at an inclination angle $\lesssim 45\degr$.

% describe physical model for flattened case
For the density structure of the flattened model, we take an ellipsoidal adaptation of the spherical description from Eq.~(\ref{eq:n1D}), where two of the three spatial axes are identical to the spherical case and the third axis is compressed. The spherical density structure is first translated into cylindrical coordinates $(R, z)$ and subsequently flattened by compressing the $z$-direction by a factor of $3$. The density scaling factor $n_0$ is scaled up by the same factor, such that the total molecular mass of the model is conserved. The temperature structure in the cloud is taken to depend on the cumulative column density toward the central position, which is extracted from the one-dimensional case (Sect.~\ref{sec:static1Dmodels}). There is no bulk velocity structure in this model. The turbulent velocity is constant at 2\,\kms, as in the spherical static and spherical infalling models from Sects.~\ref{sec:static1Dmodels} and \ref{sec:dynamicmodels}. For the radiative transfer, we use the axisymmetric version of the {\sc ratran} code \citep{hogerheijde2000}.

% ------ Figure: 2D model panel for HCO+ ------
\begin{figure}
  \centering
  \resizebox{\hsize}{!}{\includegraphics{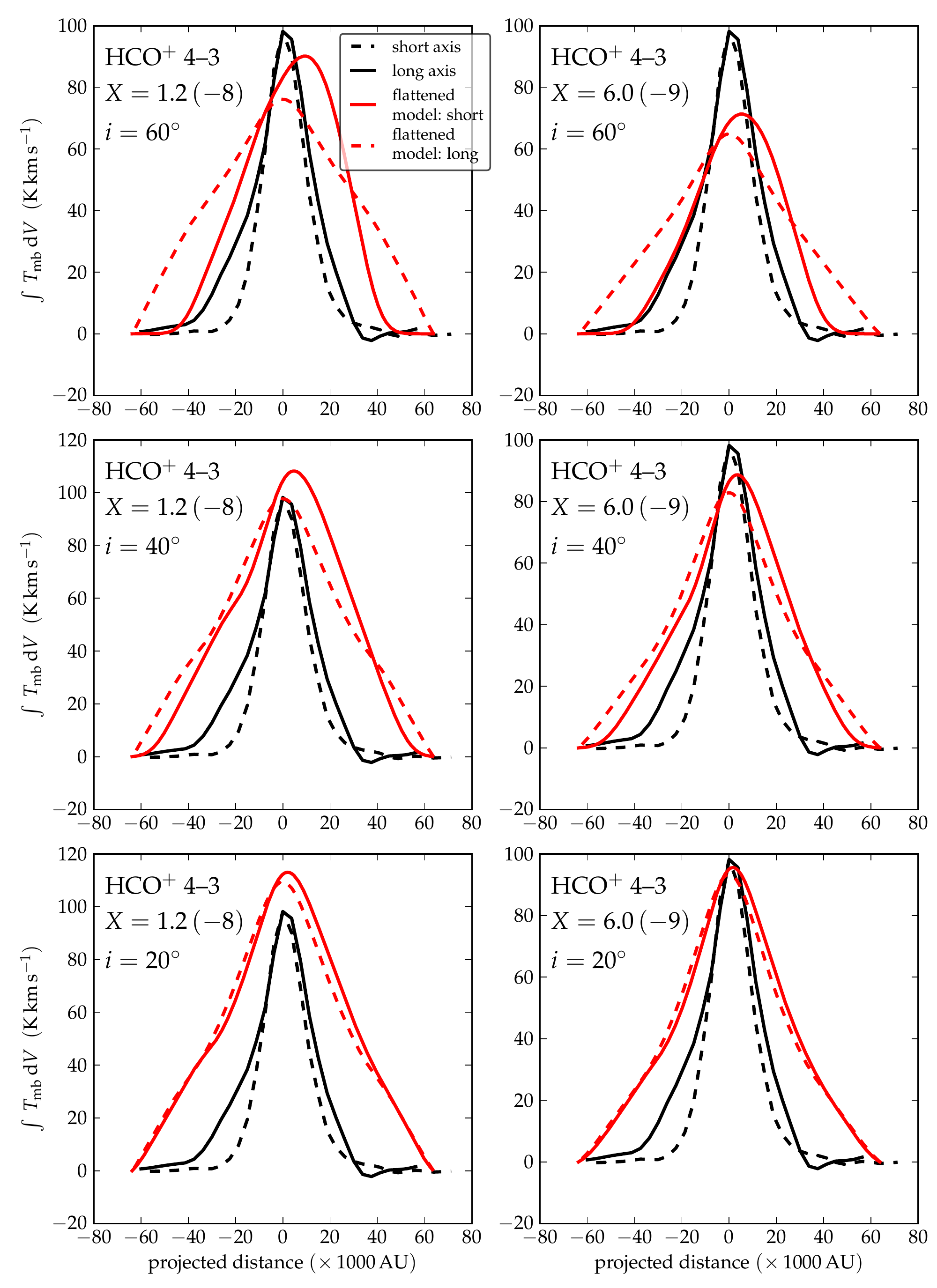}} 
  \caption{
  Position dependence of integrated line intensities for \HCOplus~\mbox{4--3}, compared with integrated line intensity profiles from flattened ellipsoidal physical models (Sect.~\ref{sec:flatmodels}). Model slices are shown for an \HCOplus\ abundance of \pow{1.2}{-8} (left column) and \pow{6}{-9} (right column). The viewing angle $i$ varies from 60\degr\ (top) to 40\degr\ (middle) to 20\degr\ (bottom). Values of optical depth at line center through the center of the model, as determined by the ray tracing routine, are for $X=$~\pow{1.2}{-8}: 42, 28, 23 at $i=60, 40, 20\degr$, respectively; and for $X=$~\pow{6}{-9}: 23, 15, 13 at $i=60, 40, 20\degr$, respectively.
    }
  \label{fig:flatHCOplusmodels}
\end{figure}

% results of flat HCO+ model
Figure~\ref{fig:flatHCOplusmodels} presents the result of radiative transfer using the ellipsoidally flattened model for \HCOplus~4--3. We choose two relative molecular abundances, $X=$~\pow{1.2}{-8} and \pow{6.0}{-9}, and three inclination angles for the ray tracing, $i=20, 40,$ and $60\degr$, where $0\degr$ is `face-on' (viewing along the $z$-axis) and $90\degr$ is `edge-on'. The best match is given by $X=$~\pow{6}{-9}, $i=20\degr$ (Fig.~\ref{fig:flatHCOplusmodels}, bottom right panel). The figure shows that in all cases, the modeled \HCOplus~4--3 distribution is much more extended than what is observed. We note that the flattened envelope at $X=$~\pow{6}{-9} and $i=20\degr$ does explain the observed \intintens\ of \HCOplus~4--3 at the central position, with a molecular abundance of \pow{6}{-9}, whereas the spherical models (Sect.~\ref{sec:static1Dmodels}, \ref{sec:dynamicmodels}) underpredict the observed \intintens\ at the central position, even with more than double the abundance ($X=$\pow{1.5}{-8}). The flattened model also has a significantly lower optical depth of 13 ($X$$=$\pow{6}{-9}, $i$$=$$20\degr$) at line center through the center of the model, where the spherical models have $\tau>20$ for \HCOplus~4--3. For $i=40\degr$, at the same low abundance, the optical depth is 15, and for $i=60\degr$ it is 23, comparable to opacities in the spherical models. 

% observed velocity-profile of HCO+ vs. flattened model
The modeled \HCOplus~4--3 line shape in velocity space is also discrepant w.r.t.~the observed line profile, which has a roughly Gaussian shape with one single peak centered at the known source velocity. Although an optical depth of 13 at line center is lower than in the spherical models, Fig.~\ref{fig:HCOplus_vprofiles} shows that the flattened model still results in a very thick line profile, where most of the photons find a way out of the cloud at $>2$\,\kms\ away from the line center. As seen before, the modeled 4--3 line of the isotopic variant \HthCOplus\ is optically thin ($\tau=0.3$), and therefore produces a better match to the observed line profile (Fig.~\ref{fig:HCOplus_vprofiles}, bottom panel).  

% test case of flattening factor 6
To test how sensitive the emerging integrated intensity profiles are to the adopted flattening factor of $3$, we also investigate the \intintens\ map of \HCOplus~4--3 resulting from a model with a flattening factor of $6$. The peak observed \intintens\ value is now best matched with an \HCOplus\ abundance of $\sim$\pow{2}{-9} (at $i=20\degr$), with an optical depth at line center of $\sim$$5$. As previously, the spatial profile of \intintens\ is too shallow with respect to the observed map. Hence, model geometries with flattening factors of $3$ and $6$ both fail to explain the observed intensity distribution of \HCOplus~4--3. While we consider flattening factors of a few to be reasonable, much flatter models would result in a geometrical description of the envelope that resembles a sheet, for which no supporting evidence exists.

% --------- DISCUSSION --------
\section{Discussion}
\label{sec:discussion}

% ------- how adequate are the RATRAN models? ------
\subsection{Physical structure of the modeled envelope}
\label{sec:discussmodels}

% recapitulate modeling approach
In Sect.~\ref{sec:modeling} we have compared the observed spatial distribution of a selection of molecular tracers with spherical static and infalling models, and with a static ellipsoidally flattened model.  
% 1D static: rare isotopes OK, match goes bad as soon as optical depth kicks in
The spectral lines of the rare isotopic variants \CtfS\ and \HthCOplus\ can be explained by the static spherical model from Sect.~\ref{sec:static1Dmodels}, both in terms of integrated intensity at emission center and shape of the spatial profile. But for the main isotopic species HCN, \HCOplus, and CS, as well as for \CstO, this representation of the molecular envelope fails to reproduce the observed integrated intensity levels. We conclude that for HCN, \HCOplus\ and CS the line optical depths of $\geq 10$ (see Fig.~\ref{fig:static1Dmodels}) are preventing radiation from escaping the model envelope. Attempts to compensate for the missing intensity by increasing the molecular abundance logically result in even higher line opacities and do not result in more photons escaping the cloud. Moreover, the already self-absorbed spectral line profiles become more discrepant w.r.t.~the observations as the abundance is increased.

\begin{figure}
  \centering
  \resizebox{\hsize}{!}{\includegraphics{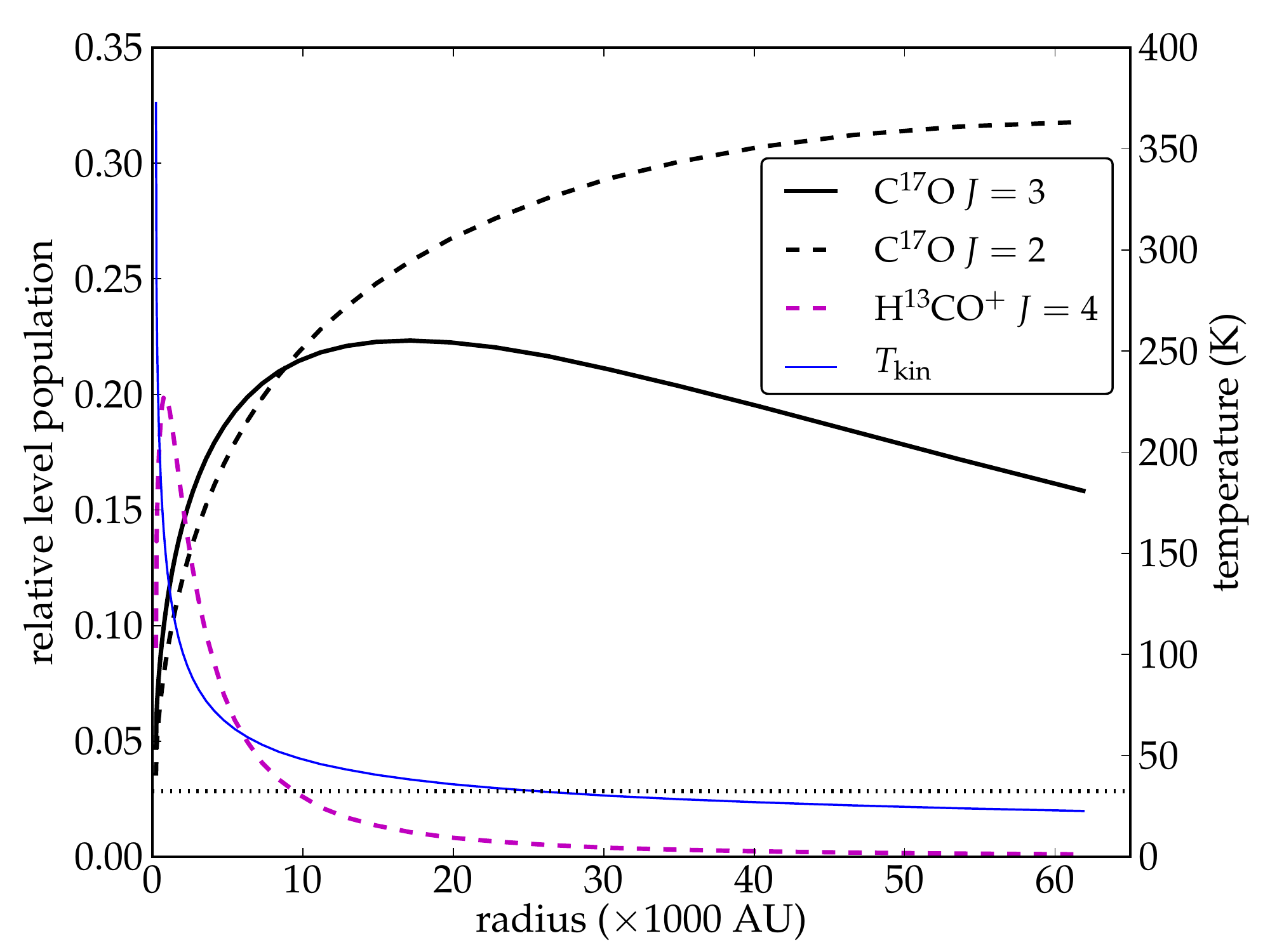}}
  \caption{Relative level populations of the $J=3$ and $J=2$ levels of \CstO, and the $J=4$ level of \HthCOplus, as determined by the Monte Carlo method in {\sc ratran} for the static spherical model. The population in each $J$ level is given relative to the total population for that species. The gas temperature profile ($T_\mathrm{kin}$) of the model cloud is shown by the solid blue line. The energy level of the $J=3$ rotational state of \CstO\ is indicated by the dotted horizontal line. }
  \label{fig:C17Olevelpops}
\end{figure}

% implication of C17O line wings, and excitation effect
% originally from section 3:
Unlike the other two lines that show a large discrepancy in spatial distribution of the emission (HCN~4--3 and \HCOplus~4--3), the modeled \CstO~3--2 emission is optically \emph{thin} ($\tau = 0.1$) at line center through the center of the model cloud. The low optical depth is confirmed by inspection of modeled line profiles in velocity space, which are well represented by a Gaussian, both on and off center. We suspect that the relative overproduction of \CstO~3--2 away from the central position is caused by excitation effects. Figure~\ref{fig:C17Olevelpops} shows that the level populations of the $J=3$ level determined by the Monte Carlo method peaks at 18\,000\,AU (outside the FWHM of the central beam of $15\arcsec \equiv 15\,000$\,AU), and declines steeply toward smaller radii. If higher $J$-levels are examined, it becomes apparent that the high kinetic temperatures in the center of the envelope leaves lower levels such as $J=3$, with an energy of only $32$\,K above ground, sparsely populated. 

The mismatch with the observed spatial distribution of \CstO\ could imply that either the description of the excitation conditions (i.e., temperature) is inadequate, or an additional component of cold, tenuous gas should be added to the model. The current truncation radius of $64$\,kAU, at which point the temperature is $22.6$\,K, is derived from CS line observations by \citet{vandertak2000jul}. An extension of the model envelope beyond this radius, to lower densities and lower temperatures, would contribute to a cold, tenuous component. The latter explanation is favored, since adjusting the temperature profile would affect many molecular species, whereas adding low-density material would mostly affect CO and its isotopes. However, two test models with extended envelopes, following the same density slope as in Sect.~\ref{sec:static1Dmodels} and with a temperature extrapolated following $r^{-0.4}$, suggest that the truncation radius alone cannot explain the discrepancy. We define one model with additional shells extending to $126$\,kAU and a temperature ranging down to $17.0$\,K, and another extending to $221$\,kAU and a temperature of $13.6$\,K. The spatial distribution of integrated \CstO~3--2 intensity resulting from the first model is steeper, but still too flat to match the observed distribution. The \intintens\ profile resulting from the second extended model is very similar to the first one, suggesting that a further extension of the envelope would not bring the modeled intensity distribution closer to the observations.
Furthermore, we point out that by incorporating the observed blueshifted component of, a.o., \CstO~3--2 emission we could be contaminating our view of the quiescent molecular envelope. 

% infalling models inadequate for thick lines
An attempt to reduce line opacity by introducing velocity structure (Sect.~\ref{sec:dynamicmodels}) is moderately successful in the sense that line optical depths are indeed decreased, but not sufficiently to explain the spatial distribution of integrated line intensity for species such as \HCOplus\ and HCN.  
Moreover, the fact that the spectral linewidths produced by the infall model are incompatible with observations of both optically thick as well as optically thin lines (Fig.~\ref{fig:HCOplus_vprofiles}), could mean that, if any velocity gradient exists in the molecular envelope, complete free-fall is not the correct approach. Part of the envelope may be radiatively, magnetically or turbulently supported against collapse, while other (outer) parts are in fact still collapsing. 

% discuss poor match of flattened HCO+ in sky plane
Finally, the flattening of the spherical cloud in one spatial direction (Sect.~\ref{sec:flatmodels}) also reduces optical depths, but only when viewed under a moderate inclination angle ($\sim$$20\degr$, see Fig.~\ref{fig:flatHCOplusmodels}). A high inclination angle has the additional effect of producing spatial profiles which are more asymmetric than what is observed. Hence, if the envelope of AFGL2591 is flattened in reality, we expect the system to be viewed almost face-on. We note that both trial \HCOplus\ abundance values in Sect.~\ref{sec:flatmodels} are consistent with \pow{1}{-8} to within the uncertainty margin of a factor $2$ quoted by \citet{vandertak1999}. 

% discuss (mis)match of flattened HCO+ in velocity space
In addition, it is apparent from Fig.~\ref{fig:HCOplus_vprofiles} that neither the infall model nor the flattened model result in velocity profiles of the emission lines which are as narrow as the observed linewidths. We note that the discrepancy of the velocity profile for both the optically thin and thick lines is larger for the infall model than for the static flattened model. It is an interesting result that the change from static spherical to static ellipsoidal leaves the line shape of the optically thin \HthCOplus~4--3 intact, while it also manages to let more of the optically thick H$^{12}$CO$^+$~4--3 line intensity escape the cloud. 

% HCN specific: abundance jump? high critical density
For HCN specifically, \citet{boonman2001} and \citet{stauber2005} have suggested an abundance which is enhanced by as much as a factor 100 in the central few hundred AU. We introduce an abundance jump from \pow{1}{-8} at $T_\mathrm{gas}<230$\,K to \pow{1}{-6} above $230$\,K in the framework of our static spherical model (Sect.~\ref{sec:static1Dmodels}), which originally has a constant abundance. The threshold for the abundance jump is motivated by gas-phase chemistry \citep{boonman2001,vandertak1999}, rather than by sublimation from grain surfaces, which puts the evaporation threshold at $\sim$100\,K for H$_2$O ices and at $\sim$25\,K for CO ices. While this abundance jump is in principle a candidate to explain the missing line intensity from the central region of the envelope (cf.~Fig.~\ref{fig:static1Dmodels}), the abundance jump model underproduces the observed HCN~4--3 intensity toward the central position by $\sim$$40\%$, while simultaneously overproducing the off-center intensity. This is caused by the fact that only the central few hundred AU enjoys a gas temperature above 230\,K, the effects of which are washed out by the $\sim$$15\arcsec$ (15\,000\,AU) FWHM of the telescope beam. Hence, the very inner part of the envelope affected by the enhanced abundance is too small to modify the \intintens\ profile on spatial scales probed by the SLS observations. 
Alternatively, the lack of modeled HCN~4--3 from the center of the envelope can be explained (partly) by the critical density, which is on the order of $10^8$\,\pccm\ for HCN~4--3 (at $100$\,K) and only $\sim$$10^7$\,\pccm\ for \HCOplus~4--3. Compared to the maximum density of \pow{1.2}{7}\,\pccm\ in the central shell of the spherical models (Sects.~\ref{sec:static1Dmodels}, \ref{sec:dynamicmodels}), it is found that HCN is not thermalized. 

% discuss results from dust continuum models
The dust continuum modeling in Sect.~\ref{sec:dustymodels} indicates that a spherical description of the envelope with a single power law density profile cannot simultaneously explain the 450\,\micron\ and the 850\,\micron\ continuum maps. This could indicate a deviation from spherical symmetry. 
%  and steeper density test models
Test runs of line radiative transfer models with density indices steeper than $1.0$ (Sect.~\ref{sec:diffalpha}) demonstrate that model descriptions with $\alpha=1.5$ or $2.0$ do not provide a more satisfying match to the observed emission in spatial and spectral dimensions simultaneously. 
% connect our envelope to other high-mass and low-mass envelopes from literature
Other envelopes around young high-mass stars are generally well fit by power law density profiles with $\alpha$ between $1.0$ and $2.0$, as studied in the samples by \citet{vandertak2000jul} and \citet{hatchell_vdtak2003}. Density slopes in low-mass protostellar envelopes are found to occupy the same range \citep{jorgensen2002}. The slope of $\alpha=1.0$ for AFGL2591 is on the shallow end of this interval, but a direct correlation between $\alpha$ and evolutionary stage is not found in comparitive studies by, e.g., \citet{vandertak2000jul}.

% concluding: so what can be the ultimate solution?
Concluding, the model descriptions employed in this study only match the observed molecular line intensity distributions if the line under consideration is optically thin. From a modeling point of view, signatures from these lines are not sensitive to the detailed distribution of the material in the envelope, as long as roughly the right amount of material is present at the right density and temperature. Conversely, we find a profound mismatch for optically thick lines, exactly those that are observational probes of the geometry of the envelope and the distribution of the material. 
% propose solutions:
Below we propose several solutions to the shortcomings of our models. 
\begin{itemize}
	\item Instead of choosing between either velocity structure or non-spherical morphology, a \emph{combination} of these two ingredients might yield a closer match of the observed molecular emission in the spatial domain. In the spectral domain, however, we do not expect a non-spherical infalling model to exhibit line shapes that will match the observations (cf.~Fig.~\ref{fig:HCOplus_vprofiles}). 
	\item A second route to lower optical depths is the adoption of an outflow cavity in the envelope, in combination with a favorable viewing angle, as proposed by \citet{vandertak1999} and later investigated by \citet{preibisch2003} and \citet{bruderer2010a}. 
	\item A third option is to introduce inhomogeneous (`clumpy') structure to alleviate optical depth effects along selected lines of sight \citep{wang1993,spaans_vandishoeck1997}. The spatial scales of these inhomogeneities should then be such that they are largely unresolved by the observations presented here, i.e., smaller than $\sim$$10^4$\,AU. 
\end{itemize}

% ------ discuss conclusions from non-modeled maps ------
\subsection{Further evidence of substructure}
\label{sec:nonmodeled}

% introduction of phenomenological discussion
The models described in Sect.~\ref{sec:modeling} are an attempt to explain the large-scale morphology of the molecular envelope of AFGL2591 by focusing on a selection of molecular species, but we have also noted additional irregular structure in other tracers in Sects.~\ref{sec:results_distr} and \ref{sec:results_velo}. Various maps of molecular species which we did not model explicitly are therefore interesting to approach from a phenomenological perspective. 

% four-directional outflow?
The northeastern `arm' together with the southern indentation seen in six different species might indicate a quadrupolar outflow. This could be interpreted as a set of two bipolar outflows originating from two separate protostars, instead of the previously assumed bipolar outflow associated with a single protostar. However, apart from the velocity separation seen in \NtwoHplus, there is no overwhelming kinematic evidence for this additional outflow direction.
Alternatively, the `arms' at position angles $\sim$130\degr\ and $\sim$310\degr\ (\NtwoHplus\ velocity map in Fig.~\ref{fig:velmapsCO_N_HxCO}) may also be \emph{infalling} gas streams, perhaps belonging to a larger scale filamentary structure \citep{klessen2005}. 

% NE arm bright in methanol, not seen in CCH nor in S-species
The northeastern plume is especially bright in several \methanol\ transitions with varying $E_\mathrm{up}$. This is an indication that its origin does not lie in excitation effects. Combined with the observation that the plume is absent in \CCH\ and the sulfur-bearing species, we suggest that chemical effects are at play. Grain surface chemistry is known to be an important formation route for \methanol\ \citep{charnley1997,herbst_vandishoeck2009}, which could have evaporated from grain mantles under the influence of a local shock front. 

The shape and direction (position angle $\sim$$230\degr$) of blueshifted components of \thCO, \CstO, \HCOplus, o-\HHCO, and \NtwoHplus\ coincide with the direction of the known approaching molecular outflow \citep{hasegawa1995}. It is interesting to note that, contrary to most molecules in Figs.~\ref{fig:velmapsCO_N_HxCO} and \ref{fig:velmapsS_CH3OH_C2H} where both the envelope and the outflow contribute to the observed emission, \NtwoHplus~4--3 in particular appears to trace the outflow rather than the bulk envelope. We base this on the significantly flattened morphology of its integrated line emission map as well as the velocity gradient which is more pronounced for \NtwoHplus\ than for the other molecules observed here. The \NtwoHplus\ molecule has previously been used as a tracer of dense material \citep[e.g.,][]{caselli2002c,caselli2002a,crapsi2004,difrancesco2004}, but usually from observations of the $J$$=$$1$--$0$ or 3--2 transition. We note that maps of \NtwoHplus~4--3 are rare in the literature. Moreover, \citet{busquet2011} have found significant depletion of \NtwoHplus\ w.r.t. NH$_3$ in dense parts of hot cores, again derived from \NtwoHplus~1--0 observations. So apart from the pronounced velocity structure, density dependent depletion could contribute to the atypical distribution of \NtwoHplus\ emission seen in our map (Fig.~\ref{fig:mapsCO_N_HCOplus}). 
Conversely, the \NtwoHplus\ emission could trace the evaporation of N$_2$ \citep{bergin2002a} along outflow cavity walls.

% suggested explanations for peak position offsets
Although only marginally significant w.r.t.~the pointing uncertainties, the fact that high-density tracers such as HNC and CN have peak positions offset to the south w.r.t.~other species could be a result of density anisotropies. \CCH\ being offset to the northwest could arise from chemical processes influenced by non-isotropic UV-radiation. 
%Similarly, the flattened morphology of warm \SOtwo\ perpendicular to the main outflow directions indicates inhomogeneous structure of the envelope. 
%A peak position offset of SO (Sect.~ \ref{sec:results_distr}) toward the receding outflow direction is recognized, possibly caused by the relatively large contribution of redshifted emission to the total \intintens\ for SO (Sect.~\ref{sec:results_velo}). On the other hand, the distribution of blue and red emission for SO, as well as for \SOtwo, is spatially more coincident than for other species. If \SOtwo\ and SO, which show no elongation on the sky, is to be interpreted as tracing any velocity structure, then it is primarily velocity structure along the line of sight (? infall, or outflow?).

% separated warm methanol peaks
The warm \methanol\ $12_1$--$12_0~A^\mp$ transition, with $E_\mathrm{up}$$=$197\,K, shows a two-peak structure (see Sect.~\ref{sec:results_spatialextent}), indicative of two separated heating sources. While four emission sources with separations $\lesssim6$\arcsec\ have been seen in interferometric observations of this source \citep{vandertak1999,trinidad2003,benz2007}, the \emph{warm} components separated by $>10$\arcsec\ ($>$10\,000\,AU) described here have not been seen before. Although the $12_1$--$12_0$ transition is not listed in studies of masing \methanol\ lines \citep[e.g.,][]{cragg1992}, we cannot exclude the possibility that it is a maser. If it is, then a linewidth narrower than the $5$\,\kms\ that we observe would need to be confirmed by higher angular resolution observations. An intrinsically narrow line profile is potentially smeared out by our $15$\arcsec\ beam, which encompasses emission from a considerable portion of the molecular cloud. Moreover, excitation analysis of an ensemble of \methanol\ lines, planned in \tSLSchempaper\ will reveal whether or not the $12_1$--$12_0$ emission fits in the picture of thermal emission.

% ------- conclusion/summary and outlook -------
\subsection{Conclusions and outlook}

% ----- overview cartoon
\begin{figure}
  \centering
  \resizebox{\hsize}{!}{\includegraphics[angle=0]{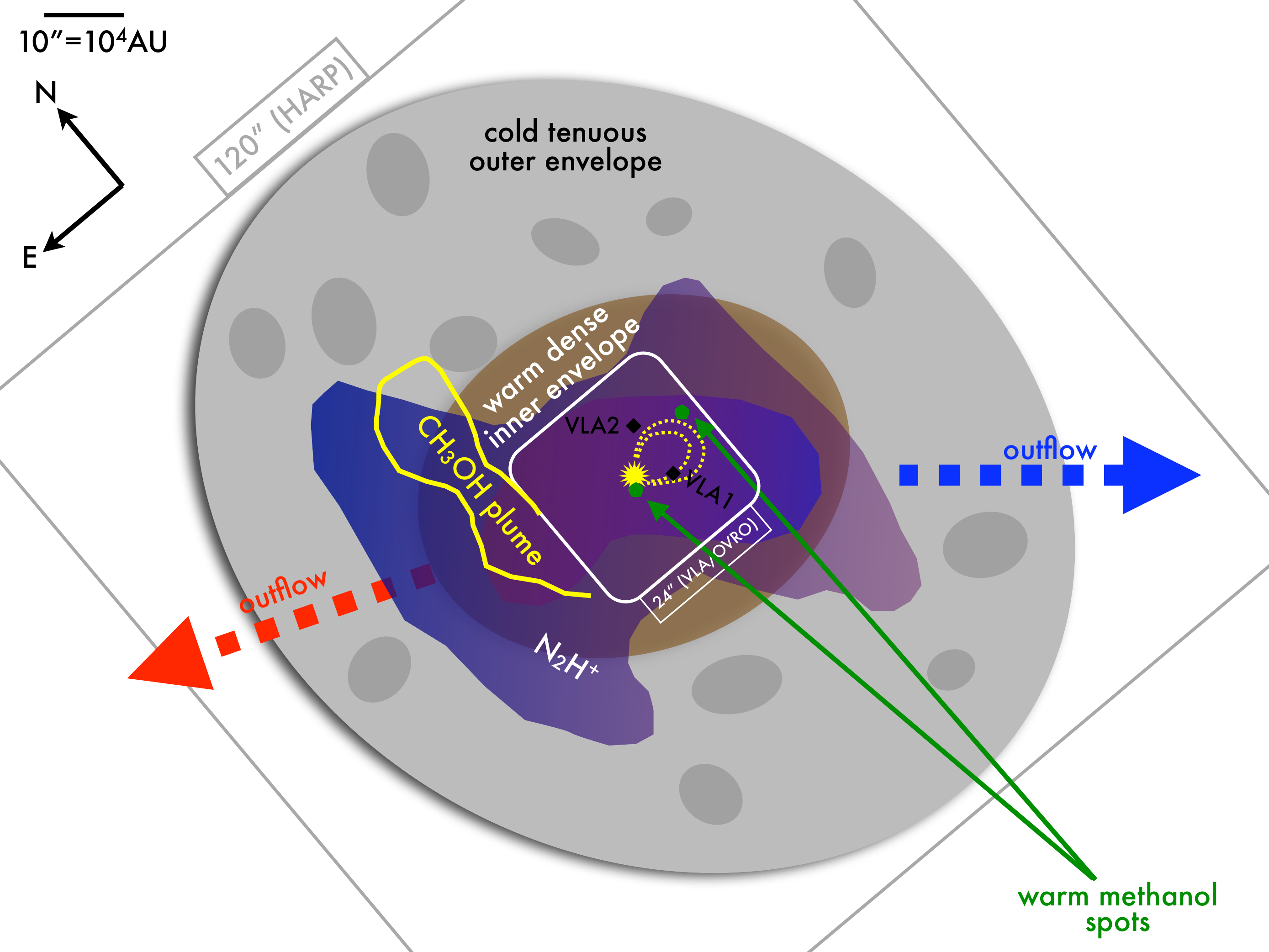}}
  \caption{Overview of observed substructure in the circumstellar envelope of AFGL2591. The `outer' and `inner' envelopes, as well as the four-arm structure of \NtwoHplus, the warm \methanol\ spots, and the \methanol\ plume are described in this work. The directions of the approaching (blue) and receding (red) outflows are indicated. The yellow star marks the position of the central heating source (infrared source from \citealt{vandertak1999}). VLA1 and VLA2 are radio continuum sources detected by \citet{trinidad2003}. The near-IR loops from \citet{preibisch2003} are indicated in dashed yellow. For reference, the fields of view of JCMT/HARP-B and the interferometric maps from VLA and OVRO are indicated. }
  \label{fig:structurescales}
\end{figure}

To summarize, this study uses the unbiased spectral imaging from the JCMT Spectral Legacy Survey to investigate the structure of the protostellar envelope of AFGL2591. We employ 35 spatially resolved molecular line signatures to investigate the envelope in three dimensions: two spatial axes with a resolution element of $\sim10^4$\,AU in projected distance, and the line-of-sight velocity axis with a resolution of 1\,\kms. 

Figure~\ref{fig:structurescales} provides an overview of selected observations from the literature and from this study which are indicative of substructure in the envelope of AFGL2591. With this overview we aim to support the suggestion that inhomogeneous substructure at scales of a few 1000\,AU and smaller should be present in molecular envelopes around protostars. 
% argument for clumpy structure: double-peak methanol, flattened warm SO2, indent/arm-structures in other species
The two separated spots of warm \methanol, a warped asymmetric plume feature, a flattened morphology of warm \SOtwo, and the multiple `arm' structures seen in CO, HCN, \HHCO, \NtwoHplus, and CS (Sect.~\ref{sec:nonmodeled}) could arise from local underdense patches with associated lower cumulative column densities, higher (UV)-flux conditions, and consequently higher gas temperatures and altered chemistry. At the same time, such a picture could resolve optical depth effects for the quiescent envelope tracers discussed in Sects.~\ref{sec:modeling} and \ref{sec:discussmodels}.
% warrant further theoretical efforts
Therefore, our results warrant further theoretical modeling efforts for protostellar envelopes in the direction of clumpy models and of flattened model structures, possibly including cavities and velocity structure.
% promote scientific value of SLS
We emphasize the value of the spectral imaging survey provided by the SLS, which allows for the simultaneous employment of several molecular tracers to characterize the physical properties of the large-scale envelope of AFGL2591.

% future observations: get more maps of high-T tracers at 0.05-0.5 pc scales
An unbiased {\it Herschel}/HIFI spectral survey of AFGL2591 in the 490--1240\,GHz regime is currently under analysis \pCHESSAFGL. This will give insights into the excitation conditions in the molecular envelope, but does not contain any spatial information. 
Moreover, spectral mapping of a consistent set of molecular tracers such as those presented here, but at higher angular resolution, will be needed to confirm the `clumpy' picture sketched above. The set of tracers should be chosen to be sensitive to the overall quiescent envelope as well as to pockets of warm ($>200$\,K) actively heated gas. Although our specific target is well into the northern sky, the obvious candidate observatory to study similar high-mass star-forming envelopes in the southern sky is the Atacama Large Millimeter and Submillimeter Array\footnote{http://www.almaobservatory.org}. This facility will be able to map high-$J$ lines of simple molecules like CO, HCN, \HCOplus\ -- and their rarer isotopic variants -- at subarcsecond angular resolution.

\acknowledgements{
% referee and editor
The authors are grateful to the anonymous referee whose constructive comments helped to improve the manuscript, and to the editor, Malcolm Walmsley, for additional suggestions. 
% JCMT staff 
We acknowledge the JCMT staff for their support. 
% CADC
This research uses the facilities of the Canadian Astronomy Data Centre operated by the National Research Council of Canada with the support of the Canadian Space Agency.
}

% ---------- REFERENCES ----------
\bibliographystyle{aa}  % file aa.bst
\bibliography{../../../literature/allreferences}

\end{document}